\documentclass[a4paper,11pt]{article}
\pdfoutput=1 

\usepackage{jcappub}
\usepackage[utf8]{inputenc}
\usepackage[format=hang,font=small,textfont=it]{caption}
\allowdisplaybreaks[2]
\usepackage{xcolor}

\title{Scalar and Tensor Perturbations in DHOST Bounce Cosmology}
\author[a,b,d]{Mian Zhu,} 
\author[a,c,e,]{Amara Ilyas,} 
\author[a,c,e,g,1]{Yunlong Zheng,} 
\author[a,c,e,1]{Yi-Fu Cai,\note{Corresponding author}} 
\author[f,a,c,1]{Emmanuel N. Saridakis}

\affiliation[a]{Department of Astronomy, School of Physical Sciences, University of Science and Technology of China, Hefei, Anhui 230026, China}
\affiliation[b]{Department of Physics, The Hong Kong University of Science and Technology, Clear Water Bay, Hong Kong S.A.R., China}
\affiliation[c]{CAS Key Laboratory for Researches in Galaxies and Cosmology, University of Science and Technology of China, Hefei, Anhui 230026, China}
\affiliation[d]{HKUST Jockey Club Institute for Advanced Study, The Hong Kong University of Science and Technology, Clear Water Bay, Hong Kong S.A.R., China}
\affiliation[e]{School of Astronomy and Space Science, University of Science and Technology of China, Hefei, Anhui 230026, China}
\affiliation[f]{National Observatory of Athens, Lofos Nymfon, 11852 Athens,
Greece}
\affiliation[g]{ICRANet, Piazza della Repubblica 10, I-65122 Pescara, Italy} 

\emailAdd{mzhuan@connect.ust.hk}
\emailAdd{aarks@mail.ustc.edu.cn}
\emailAdd{zhyunl@ustc.edu.cn}
\emailAdd{yifucai@ustc.edu.cn}
\emailAdd{msaridak@phys.uoa.gr}

\date{}

\abstract{We investigate the bounce realization in the framework of DHOST 
cosmology, focusing on the relation with  observables. We 
perform a detailed analysis  of the scalar and tensor perturbations during the 
Ekpyrotic contraction phase, the bounce phase, and the fast-roll 
expansion phase, calculating the   power spectra,  the 
spectral indices and the tensor-to-scalar ratio. Furthermore, we study the  
  initial conditions, incorporating   perturbations generated 
by Ekpyrotic vacuum fluctuations, by matter vacuum fluctuations, and by thermal 
fluctuations.  The scale invariance of the scalar power spectrum can be acquired introducing a matter contraction phase before the Ekpyrotic phase, or invoking
a thermal gas as the source. The DHOST bounce scenario  with   cosmological perturbations  generated by  thermal fluctuations proves to be the most efficient one, and the 
corresponding predictions are in perfect agreement with   observational 
bounds. Especially the tensor-to-scalar ratio is many orders of magnitude within
the allowed region, since  it is suppressed by the   Hubble 
parameter at the beginning of the bounce phase.  
}

\begin{document}
	\maketitle
	
\section{Introduction}
\label{sec:intro}
	
	Bounce cosmology \cite{Novello:2008ra, Lehners:2008vx, Cai:2014bea, 
Battefeld:2014uga, Brandenberger:2016vhg, Cai:2016hea} can serve as a very early universe scenario for the structure formation alternative to the standard 
inflationary cosmology (see e.g. \cite{Brandenberger:2011gk} for critical reviews on 
inflation and its alternatives). Nonsingular bounce cosmology has the merit to 
avoid the initial singularity problem \cite{Borde:1993xh, Borde:2001nh} 
encountered by inflationary cosmology, which motivates the investigations on 
this field. However, bounce models also face new challenges 
such as the ghost and gradient instability problem \cite{Cline:2003gs, 
Vikman:2004dc, Xia:2007km, Easson:2016klq}, the problem of anisotropic stress 
\cite{Karouby:2010wt, Karouby:2011wj, Bhattacharya:2013ut, Cai:2013vm} (also 
known as the Belinski–Khalatnikov–Lifshitz (BKL) instability \cite{Belinsky:1970ew}), and the tension between 
tensor-to-scalar ratio and non-Gaussianities \cite{Cai:2009fn, Gao:2014hea, 
Gao:2014eaa,Quintin:2015rta, Li:2016xjb, Akama:2019qeh}. Nevertheless, it is     important to discuss the validity of 
nonsingular bounce cosmology from the viewpoint of cosmological observations.
	
Many attempts have been made to evade the aforementioned problems in
constructing a nonsingular bounce scenario \cite{Kumar:2021mgc, Pinto-Neto:2021gcl,Nandi:2020szp}. In the literature, higher-order 
scalar-tensor theories  beyond the Horndeski/Generalized Galileon theory are introduced to solve the gradient instability problem 
\cite{Cai:2016thi,Creminelli:2016zwa,Cai:2017tku,Cai:2017pga,Cai:2017dyi, 
Kolevatov:2017voe,Ye:2019frg,Ye:2019sth, Gungor:2020fce,Zheng:2017qfs}, which occurs  inevitably  in bounce 
models constructed within second-order scalar-tensor theories 
\cite{Libanov:2016kfc,Kobayashi:2016xpl, Akama:2017jsa}. Moreover, the 
Ekpyrotic scenario \cite{Khoury:2001wf}, during which the energy density of the 
dominated matter fluid blue-shifts faster than the anisotropic stress, can 
prevent the development of the BKL instability. Additionally, the tension 
between tensor-to-scalar ratio and non-Gaussianities predicted by matter bounce 
cosmology can be evaded by generalizing the action of the matter component 
\cite{Akama:2019qeh}. As a next step after addressing the conceptual issues of nonsingular bounce models, it is important to study their observational quantities, such as the amplitude and the spectral index of primordial curvature perturbations, as well as the associated tensor-to-scalar ration, so that these models can be testable in observations, and confront the predictions with    observational 
data.
	
In this work we focus on one class of nonsingular bounce cosmology, namely
the DHOST bounce developed in \cite{Ilyas:2020qja} in the framework of 
Degenerate Higher-Order Scalar-Tensor (DHOST) theories. The 
DHOST bounce models are shown to be free from ghost, gradient and 
BKL instabilities, and thus they  provide a solid realization of a healthy 
nonsingular bounce scenario. It is then natural to ask whether such models can 
survive from the observational confrontation. We shall study the scalar and 
tensor power spectra within the simplest case, namely the DHOST bounce scenario with a single bounce phase\footnote{The universe starts from a finite 
configuration and then it undergoes a contraction phase; at certain scales the 
universe stops contracting, exhibits the bounce, and transits to the expanding 
phase of Standard Big Bang cosmology. }.
	
The manuscript is organized as follows. In Section \ref{sec:review} we 
briefly review the basics of DHOST bounce, and describe the specific model that shall be analyzed. Then we investigate the background dynamics and provide its 
parametrization in Section \ref{sec:bg}. The dynamics of cosmological 
perturbation is investigated in Section \ref{sec:pt}, and the power spectra 
with different initial conditions are calculated in Section \ref{sec:initial}. 
Finally, we conclude   with relevant discussions in Section 
\ref{sec:conclusion}.
	
Throughout the work we consider the signature of the metric as $(+,-,-,-)$, and 
we denote the reduced Planck mass by $M_p \equiv 
1/\sqrt{8\pi G} = 1$, where $G$ is the Newton’s gravitational constant,
 and all used parametric values are in Planck units. The dot symbol represents 
differentiation with respect to the cosmic time $t$: $\dot{\phi} \equiv d\phi / 
dt$, and a comma in the subscript denotes  partial derivative: $\phi_{,\mu} 
\equiv \partial_{\mu} \phi$. We define $X \equiv \frac{1}{2} \nabla_{\mu} \phi \nabla^{\mu} \phi$ to be the canonical kinetic term of the scalar field. The subscript ``B-'' and ``B+'' labels the 
start/end of the bounce phase  (e.g.  $t_{B-}$ represents the 
cosmic time at the beginning of the bounce phase). Additionally, at the bounce 
point we normalize the scale factor as $a_B = 1$ and the cosmic time as $t_B = 
0$, while the conformal time, defined as $\tau \equiv \int dt/a$, is also set 
to 0 at the bounce point $\tau_B = 0$. Furthermore, we use the subscripts $f_X$ 
and $f_{\phi}$ to denote $f_X \equiv \partial f/\partial X$ and $f_{\phi} 
\equiv \partial f/ \partial \phi$, respectively. Lastly, we focus on a flat  Friedmann-Robertson-Walker (FRW) geometry with metric 
\begin{equation}
 ds^2= 
 dt^2-a^2(t)\,
\delta_{ij} dx^i dx^j, 
\end{equation}
with $a(t)$ the scale factor.

	\section{The basics of DHOST Bounce}
	\label{sec:review}
	
 In this section, we briefly review the DHOST bounce scenario developed 
in \cite{Ilyas:2020qja}.

	\subsection{The Generic Action}
	
 The generic action is taken to be
	\begin{align}
		\label{eq:Action}
		S \nonumber = \int d^4x \sqrt{-g} \Big[ & - {M_p^2}[1 + f(\phi,X)] \frac{R}{2} 
+ K(\phi,X) + Q(\phi,X) \Box \phi \\
		& - \frac{f}{4X} \left( L_1^{(2)} - L_2^{(2)} \right) + 
\frac{f-2Xf_X}{4X^2} \left( L_4^{(2)} - L_3^{(2)} \right) \Big ] ~,
	\end{align}
where $R$ represents the Ricci scalar (thus the term $-R/2$ corresponds to the 
standard Einstein-Hilbert action). Additionally, the term $K+Q\Box \phi$ is a 
part of the Galileon/Horndeski action \cite{Nicolis:2008in, 
Deffayet:2011gz,Kobayashi:2011nu,Horndeski:1974wa}, which has been introduced to 
cosmology and analyzed in 
\cite{Kobayashi:2010cm,Deffayet:2010qz,Qiu:2011cy,Easson:2011zy,Cai:2012va,
Leon:2012mt,CANTATA:2021ktz}. The function $f \equiv f(\phi,X)$ represents the 
DHOST coupling:
\begin{equation}
	\label{eq:LD}
	\mathcal{L}_D \equiv -M_p^2 f(\phi,X) \frac{R}{2} - \frac{f}{4X} \left( L_1^{(2)} - L_2^{(2)} \right) + 
	\frac{f-2Xf_X}{4X^2} \left( L_4^{(2)} - L_3^{(2)} \right) ~.
\end{equation}

We provide a brief introduction to the quadratic DHOST theory in 
Appendix \ref{app:DHOST}, and we show that the action \eqref{eq:Action} belongs 
to the $^{(2)}N-I$ type DHOST theory \footnote{ We will show in Appendix \ref{app:DHOST} that the action \eqref{eq:Action} can be classified as either the merge of Horndeski action $K+Q\Box \phi$ with type $^{(2)}N-I$ DHOST action $\mathcal{L}_D + R/2$, or the merge of Horndeski action $R/2 + K + Q\Box \phi$ with type $^{(2)}N-II$ DHOST action $\mathcal{L}_D$. As we will see in next sections, the action \eqref{eq:LD} has negligible contributions to observations, so it is useful to view our theory as the merge of Horndeski theory with type N-II DHOST theory.}, thus the model is free from the 
Ostrogradsky instabilities \cite{Ostrogradsky:1850fid} which plagues most of 
higher-order scalar-tensor theories. 
	
	\subsection{Realization of DHOST Bounce with a Single Bounce}
	
In the above construction, different choices of $f(\phi,X)$ will lead to 
different cosmological bouncing scenarios. In \cite{Ilyas:2020qja} two specific 
cases were discussed. If $f \equiv f(X)$, the DHOST coupling will not affect 
the background dynamics, and the cosmological scenario will be similar to the 
previously developed ones \cite{Cai:2012va} with a single bounce phase. On 
the other hand, if $f \equiv f(\phi)$  the bouncing scenario will be 
drastically changed, and there can exist multiple bounce \cite{An:2020lkg} phases. 
	
	As stated in the Introduction, in this work we will focus on the 
simple case where $f \equiv f(X)$, and thus there exists only one bouncing 
phase, since the results are comparable to the previous work   
\cite{Cai:2012va}. The spectra with multiple bouncing phase will be studied in 
a following project, and for interested readers  we refer to the pioneer
work \cite{Brandenberger:2017pjz} on this topic.
	
	We consider the operators in   action \eqref{eq:Action} to be
	\begin{equation}
		K(\phi,X) = {M_p^2}[1-g(\phi)]X + \beta X^2 - V(\phi) ~,~ Q(\phi,X) = \gamma X 
~,~ f(\phi,X) = { c X^2+ \frac{d}{M_p^2} X^3} ~,
	\end{equation}
	where the functions $g(\phi)$ and $V(\phi)$ are defined as
	\begin{equation}
		g(\phi) = - \frac{2g_0}{e^{-\sqrt{\frac{2}{p}}\phi} + e^{b_g 
\sqrt{\frac{2}{p}}\phi}} ~,~ V(\phi) = - \frac{2V_0}{e^{-\sqrt{\frac{2}{q}}\phi} 
+ e^{b_V \sqrt{\frac{2}{q}}\phi}} ~,
	\end{equation}
	with   $g_0$, $p$, $b_g$, $V_0$, $q$, $b_V$, $\beta$, 
$\gamma$, $c$, $d$ the model parameters. We follow the convention of 
\cite{Cai:2012va} and we consider $\phi$ to be  dimensionless, which implies 
that all parameters are dimensionless,
except for $V_0$ has dimension $[M]^4$.

In the following sections   we will use the 
parameter  values 
	\begin{align}
	\label{eq:parameter}	
		& \nonumber V_0 = 10^{-8} ~,~ g_0 = 1.1 ~,~ \beta = 5 ~,~ \gamma = 3 
\times 10^{-3} ~, \\
		& b_V = 100 ~,~ b_g = 0.5 ~,~ p = 0.01 ~,~ q = \frac{6}{31}~,
	\end{align}
	and DHOST function $f(X)=450 X^2 + 500000 X^3$
	(all dimension-full parameters are measured in Planck units)
	to numerically verify the approximations of the analytic calculation. Particularly, in Section \ref{sec:initial}, we calculate the observable parameters (scalar  amplitude $A_s$, scalar spectra index $n_s$ and tensor-to-scalar ratio $r$) and show that the predicted values are compatible with Cosmic Microwave  Background (CMB) data  with the model parameters being \eqref{eq:parameter} (see (\ref{parameterresults})).
	
\begin{figure}[ht]
		\includegraphics[width=.45\textwidth]{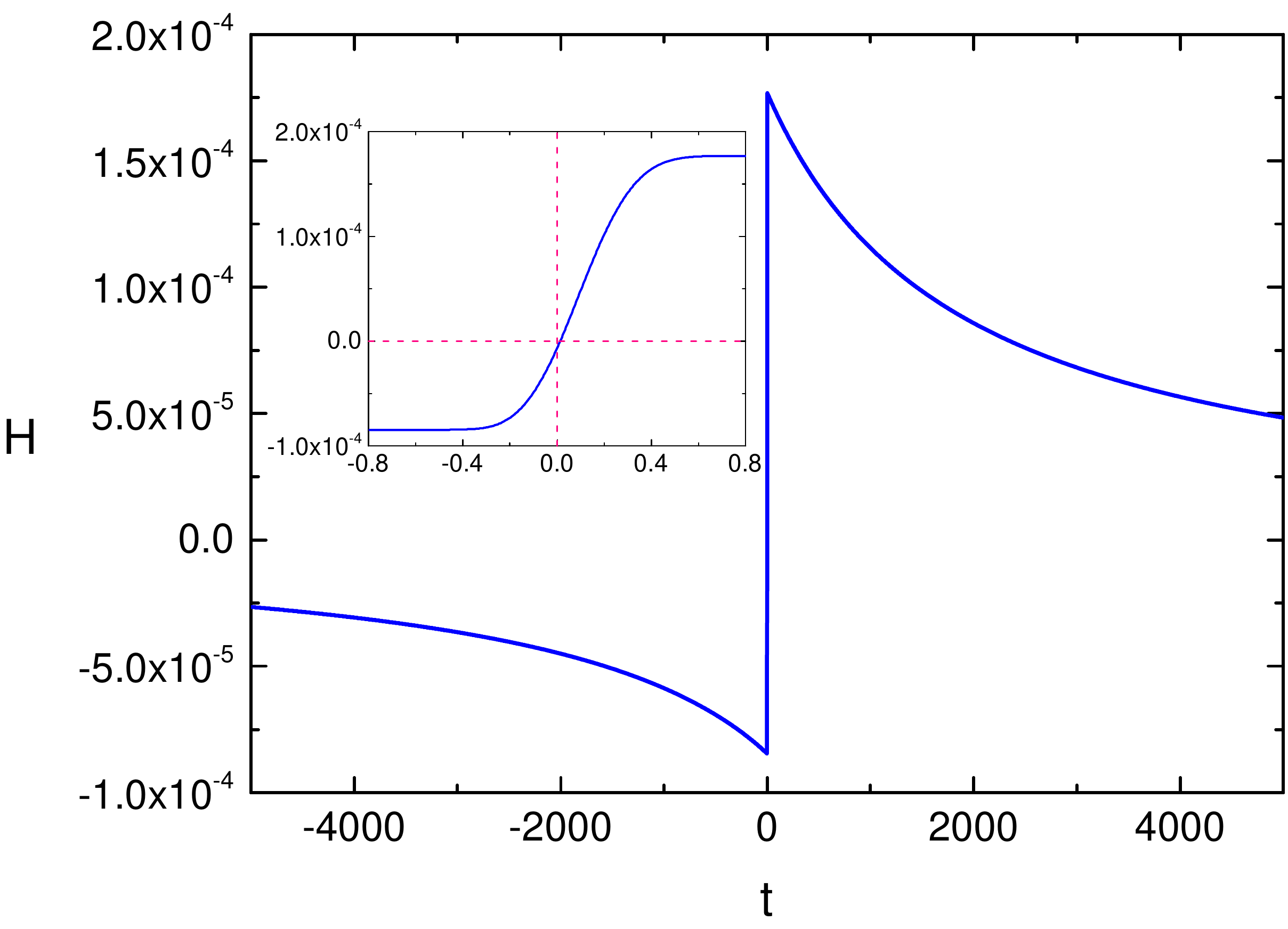} \quad 
		\includegraphics[width=.48\textwidth]{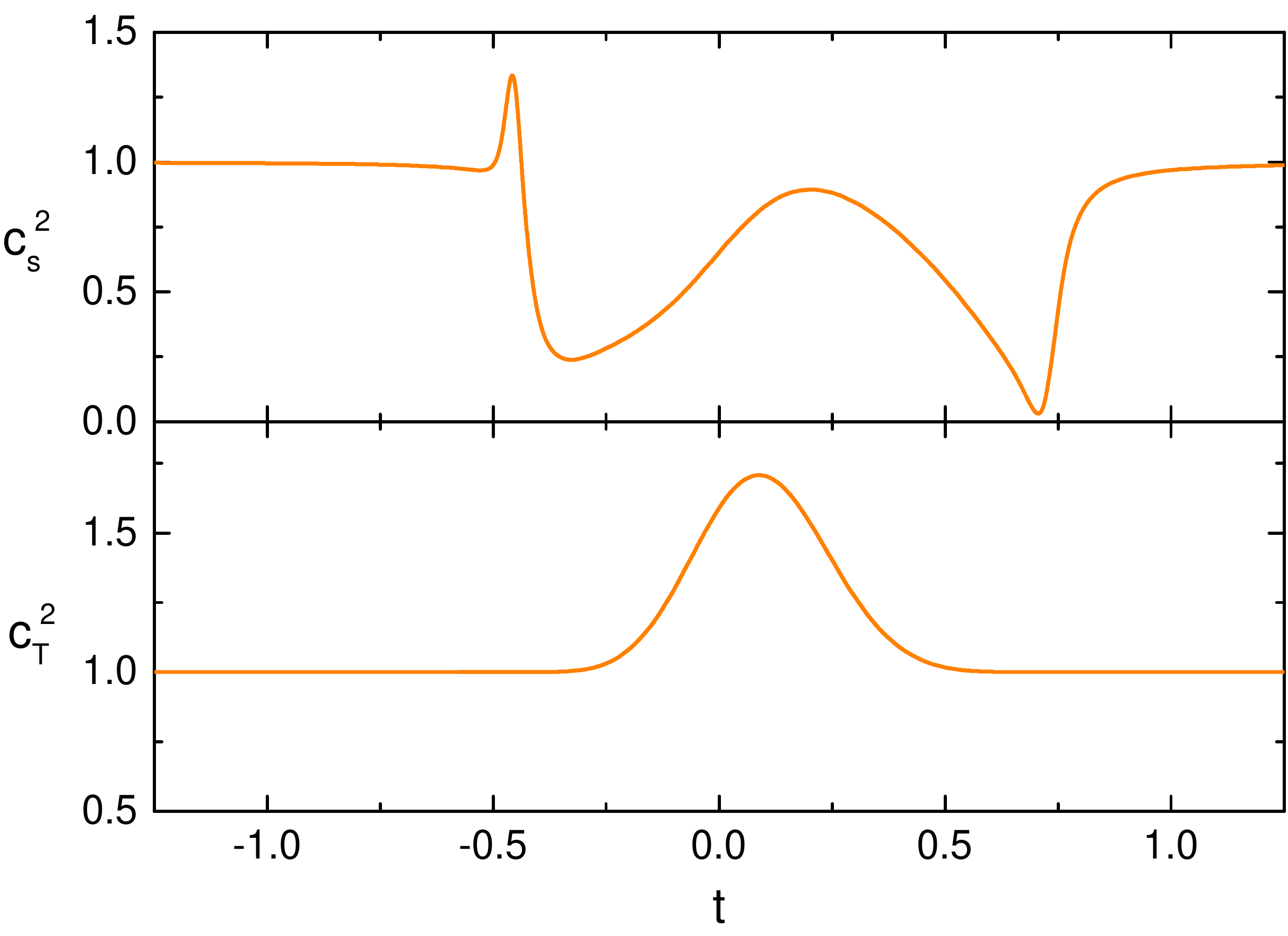} 
		\caption{Evolution of the Hubble parameter $H$ (left) and the sound 
speeds of scalar and tensor perturbations, $c_s^2$ and $c_T^2$ respectively 
(right). The background parameters are chosen 
as in  \eqref{eq:parameter}. Both $c_s^2$ and $c_T^2$ exhibit superluminal behavior near the bounce point, which is a common feature in theories beyond Horndeski \cite{Mironov:2020mfo,Mironov:2020pqh}}.
		\label{fig:Ht}
	\end{figure}

	Before proceeding, in order to provide a complete picture, in  Fig. 
\ref{fig:Ht} we depict the dynamics of the 
Hubble parameter $H\equiv\frac{\dot{a}}{a}$ and the sound speed of 
scalar/tensor perturbations $c_s^2$ and $c_T^2$ \cite{Ilyas:2020qja}. The bounce phase takes place in the neighborhood of $t=0$, where the 
Hubble parameter changes from negative to positive. Moreover, $c_s^2$ and 
$c_T^2$ are positive during the whole cosmological process, which illustrate that the model under consideration is free from the gradient instability. In the following we perform the detailed analysis, explanation, and calculation of 
observables.

	\section{Background}
	\label{sec:bg}
	
	The   background equations of motion for the model at hand are 
\cite{Ilyas:2020qja}
	\begin{equation}
		\label{eq:Friedmann1}
		3H^2 = \rho \equiv \frac{1}{2}\left[1-g(\phi)\right] \dot{\phi^2} + 
\frac{3}{4}\beta \dot{\phi}^4 + 3\gamma H \dot{\phi}^3 + V(\phi) ~,
	\end{equation}
	\begin{equation}
		\label{eq:Friedmann2}
		-2\dot{H} - 3H^2 = P \equiv \frac{1}{2}\left[1-g(\phi)\right] 
\dot{\phi^2} + \frac{3}{4}\beta \dot{\phi}^4 - \gamma \dot{\phi}^2 \ddot{\phi} - 
V(\phi) ~,
	\end{equation}
	where $\rho$ and $P$ are the energy density and pressure of the scalar 
field. Conventionally, one may define the equation of state (EoS) parameter as 
$w \equiv P/\rho$. Note that the DHOST coupling $f(X)$ is absent from  the 
equations   \eqref{eq:Friedmann1} and \eqref{eq:Friedmann2}, hence the 
background evolution is the same as in the Horndeski/Galileon bounce proposed 
in \cite{Cai:2012va}. Therefore, the whole cosmological evolution can be 
separated into three phases: the Ekpyrotic phase where the universe undergoes 
a contraction phase, the nonsingular bounce phase where the universe transits 
from contraction to expansion, and the fast-roll expansion phase which finally 
connects with the standard Big Bang cosmology. We can then follow the 
parametrization from \cite{Cai:2012va}, as described below. The numerical 
results for the evolution of the Hubble parameter are shown in 
Fig.~\ref{fig:Ht}.

	\subsection{Ekpyrotic Contraction Phase}
	
	This phase begins at the far past where $\phi \ll -1$, and ends when the 
function $g(\phi)$ reaches the critical value $1$. In this phase, when $g(\phi)$ tends to $0$, the Lagrangian approaches the canonical form: $\mathcal{L} \to 
1/2 \partial_{\mu} \phi \partial^{\mu} \phi - V(\phi)$. Accordingly, it admits the Ekpyrotic attractor solution 
	\begin{equation}
		\phi \simeq - \sqrt{\frac{q}{2}} \ln \left[ \frac{2V_0 t^2}{q(1-3q)} 
\right] ~,
	\end{equation}
	with $q$  a parameter related to the nearly constant EoS parameter during 
this phase, $w_c$, as $w_c \simeq -1 + \frac{2}{3q}$. We mention that in 
order to make the model   free from the BKL instability \cite{Belinsky:1970ew}, 
which plagues many bouncing   scenarios 
\cite{Karouby:2010wt,Karouby:2011wj}, the energy density of $\phi$ should 
increase faster than the anisotropic stress, which indicates the constraint 
$0<q<1/3$.
	
	Since during Ekpyrotic contraction the EoS parameter $w_c$ is approximately 
constant, the scale factor $a(\tau)$ evolves accordingly as
	\begin{equation}
	    \label{eq:contractionaH}
	    a(\tau) = a_{B-} \left( \frac{\tau_- - \tau}{\tau_- - \tau_{B-}} 
\right)^{\frac{q}{1-q}} ~, \mathcal{H}\simeq \frac{-q}{(1-q)(\tau_- -\tau)}  ~,
	\end{equation}
	where $\tau_-$ and $a_{B-}$ are proper integration constants and 
$\mathcal{H} \equiv \partial_{\tau} a/a$ is the conformal Hubble parameter. The 
boundary condition, i.e. $g(\phi)$ reaches $1$ at the end of Ekpyrotic 
contraction $\tau = \tau_{B-}$, gives
	\begin{equation}
		\label{eq:tau-}
		\tau_- = 
\tau_{B-} - \frac{q}{(1-q)\mathcal{H}_{B-}} ~.
	\end{equation}

Finally, we comment that in the Ekpyrotic phase, there exists only one scalar field $\phi$ whose EoS parameter satisfies $w_c > 1$, thus the principle of finite amplitude developed in \cite{Jonas:2021xkx} is not violated in our model.
	
	\subsection{Bounce Phase}
	
	As we have mentioned, we normalize the time axis by setting  $t=0$ at the 
bounce point, i.e $H(0) = 0$, while the scale factor is normalized as
$a(0) = 1$. Then, as it has been shown in \cite{Cai:2007zv, 
Cai:2008ed,Cai:2008qw,Lin:2010pf}, the evolution of the Hubble parameter 
near the bounce point can be well approximated by a linear function of cosmic 
time $H \simeq \Gamma t$. The parametrization is valid in a class of fast bounce models, and the magnitude of $\Gamma$ is usually set by the detailed 
microphysics of the bounce. 
	
	In the specific parametric choice of \eqref{eq:parameter},  
$\Gamma$ is of the order of  $\mathcal{O}(10^{-4})$, and the beginning/ending time of bounce phase, namely  $t_{B-}$ and $t_{B+}$, is of the order of 
$\mathcal{O}(1)$. Thus,  the conformal time $\tau$ is
	\begin{equation}
		\tau = \int_0^t \frac{dt^{\prime}}{a(t^{\prime})} = \int_0^t 
e^{-\frac{1}{2}\Gamma t^{\prime 2}} dt^{\prime} \simeq t - \frac{1}{6}\Gamma t^3 + \mathcal{O}(t^5) \simeq t ~.
	\end{equation}
	This implies that in the bounce phase the cosmic time $t$ and the conformal 
time $\tau$ are almost equivalent. This  is not surprising: a fast bounce 
means that the scale factor will not change significantly during  the bounce 
phase, therefore the difference between conformal and cosmic time will be 
minor.
	
	The dynamics of the scale factor and the scalar field is then
	\begin{equation}
		\label{eq:bouncepara}
		H \simeq \Gamma t ~,~ a(t) \simeq e^{\frac{1}{2}\Gamma t^2} ~,~ 
\dot{\phi} \simeq \dot{\phi}_B \exp{\left( -\frac{t^2}{T^2} \right)}  
~,
	\end{equation}
while in terms of the conformal time	
		\begin{equation}
		\label{eq:bounceparatau}
    ~\mathcal{H} \simeq \Gamma \tau ~,~ a(\tau) \simeq e^{\frac{1}{2}\Gamma 
\tau^2} 
~. 
	\end{equation}
	Here, $\dot{\phi}_B$ denotes the value of the derivative of the scalar 
field $\phi$ at the bouncing point $t=0$, which is $\dot{\phi}_B^2 \simeq 2(g_0 
- 1)/3\beta$, while the parameter $T$ is approximately one quarter of the 
duration of the bounce.
	
	\subsection{Fast-roll Expansion}
	\label{sec:bgdynamicse}

	After the bounce, the $\dot{\phi}^2$ term decreases and thus the 
  scalar field $\phi$ will cross the critical value  where the function 
$g(\phi)$ decreases to $1$, and the scalar field abandons the ghost condensate 
state. Then the Lagrangian of $\phi$   acquires the canonical form with a 
relatively flat potential, and thus the scalar field will enter a 
fast-roll state with EoS parameter $w \simeq 1$. Hence, the scale factor will 
behave as $a \propto t^{1/3}$. In particular, we have 
	\begin{equation}
		\label{eq:expansionah}
		a(t) = a_{B+} \left( \frac{t - t_+}{ t_{B+} - t_+} \right)^{1/3} ~,~ 
H(t) = \frac{1}{3(t-t_+)} ~,
	\end{equation}
	which gives the conformal time $\tau$ as
	\begin{equation}
		\label{eq:expansiontau}
		\left( \frac{\tau - \tau_+}{\tau_{B+} - \tau_+} \right)=  \left( \frac{t 
- t_+}{t_{B+} - t_+} \right)^{\frac{2}{3}} ~,
	\end{equation}
	where $\tau_+$ and $t_+$ are integration constants. 
	
	In terms of the conformal time $\tau$, the background evolution can be 
expressed   as
	\begin{equation}
		a(\tau) = a_{B+} \left( \frac{\tau - \tau_+}{\tau_{B+} - \tau_+} 
\right)^{\frac{1}{2}} ~,~ H(\tau) = \frac{1}{3(t_{B+} - t_+)} \left( \frac{\tau 
- \tau_+}{\tau_{B+} - \tau_+} \right)^{-\frac{3}{2}} ~,
	\end{equation}
	which gives the conformal Hubble parameter as 
	\begin{equation}
		\label{eq:tau+}
		\mathcal{H}(\tau) = a(\tau)H(\tau) =\frac{1}{2(\tau - \tau_+)}~.
	\end{equation}
	
	\subsection{Matching Conditions at the Background Level}
	
	The scale factor $a$, the Hubble parameter $H$, the scalar field $\phi$ and 
its first derivative $\dot{\phi}$, should be continuous at the transition 
surfaces $\tau = \tau_{B-}$ and $\tau = \tau_{B+}$. In this section,  we   
explicitly elaborate the constraints that are required  when incorporating 
  the cosmological perturbations.
	
	However, we mention that, as we will see in the next section, $\phi$ and 
$\dot{\phi}$ will not affect the dynamics of cosmological perturbations in both 
the contraction   and   expansion phases. Hence, there is no need to impose 
the continuity condition on the scalar-field content. The same argument applies 
  to the scale factor $a$ at the end of Ekpyrotic contracting phase. 
Therefore, the remaining tasks are to determine the condition for the 
continuity of $H$ at $\tau = \tau_{B+}$, $\tau = \tau_{B-}$, and the continuity 
of $a$ at $\tau = \tau_{B+}$.
	
	At the transition surface $\tau = \tau_{B-}$, the continuity of conformal 
Hubble parameter $\mathcal{H}$ gives
	\begin{equation}
		\mathcal{H}_{B-} \equiv \mathcal{H}(\tau = \tau_{B-}) = \Gamma \tau_{B-} 
~,
	\end{equation}
	which can be used to determine the value of $\tau_-$ from equation 
\eqref{eq:tau-}.
	Similarly, at the transition surface $\tau = \tau_{B+}$, the continuity of 
$a$ and $\mathcal{H}$ gives
	\begin{equation}
		a_{B+} \equiv a(\tau_{B+}) = e^{\frac{1}{2} \Gamma \tau_{B+}^2} ~,~ \ \ 
\ \ \mathcal{H}(\tau_{B+}) = \frac{a_{B+}}{3(t_{B+} - t_+)} = \Gamma \tau_{B+} 
~,
	\end{equation}
	along with the approximation $\tau_{B+} \simeq t_{B+}$ from the bounce 
phase, where    the integration constant $t_+$ is  determined by
	\begin{equation}
		{t_+ = t_{B+} - \frac{a_{B+}}{3\Gamma \tau_{B+}}}~.
	\end{equation}
	
	\section{Evolution of Perturbations}
	\label{sec:pt}
	
	In this section, we proceed to the detailed investigation of the scalar and 
tensor perturbations. Similar to the previous section, after providing the 
perturbation equations,  we will perform the analysis for the various 
phases separately.

	\subsection{Perturbations Equations}
	The actions for the scalar and tensor perturbations at linear level are 
\cite{Ilyas:2020qja,Ilyas:2020zcb,Zhu:2021ggm}
	\begin{equation}
		\label{eq:S2S}
		S_{2,S} = \int d\tau d^3x \frac{z_s^2}{2} \left[ \zeta^{\prime 2} - 
c_s^2 (\partial_i \zeta)^2 \right] ~,
	\end{equation}
	\begin{equation}
		\label{eq:S2T}
		S_{2,T} \equiv \int d\tau d^3x \frac{z_T^2}{2} \left[ 
\gamma_{ij}^{\prime 2} - c_T^2 (\nabla_k \gamma_{ij})^2 \right] ~,
	\end{equation}
	where
	\begin{equation}
		\frac{z_s^2}{2a^2} = 3 + \frac{2\left[ \dot{\phi}^2 K_X + \dot{\phi}^4 
K_{XX} - 6H^2 + 12\gamma H \dot{\phi}^3  \right]}{(\gamma \dot{\phi}^3 - 2H)^2} 
~,
	\end{equation}
	\begin{equation}
		\left( -\frac{z_s^2}{2a^2} \right) c_s^2 = 1 + f + \frac{2}{a} 
\frac{d}{dt} \left[ \frac{a(f_X \dot{\phi}^2 - f - 1)}{2H - \gamma \dot{\phi}^3} 
\right] ~,
	\end{equation}
	\begin{equation}
		z_T^2 = \frac{1}{4}a^2 ~,~ c_T^2 = 1+f ~.
	\end{equation}	
	Varying \eqref{eq:S2S} with respect to $\zeta$ leads to the 
Mukhanov-Sasaki (MS) equation \cite{Sasaki:1983kd, 
Kodama:1985bj,Mukhanov:1988jd} for scalar perturbations, namely
	\begin{equation}
		\label{eq:MSeqscalar}
		v_k^{\prime \prime} + \left( c_s^2k^2 - \frac{z_s^{\prime \prime}}{z_s} 
\right) v_k = 0 ~,
	\end{equation}
	where $v_k \equiv z_s \zeta_k$ is the MS variable. On the other hand, the 
tensor perturbations contain two   modes. As usual we 
decompose the tensor perturbations $\gamma_{ij}$ as
	\begin{equation}
		\gamma_{ij}(\tau,\vec{x}) = \gamma_+(\tau,\vec{x})e_{ij}^+ + 
\gamma_{\times}(\tau,\vec{x})e_{ij}^{\times} ~,
	\end{equation}
	with two fixed polarization tensors, $e_{ij}^+$ and $e_{ij}^{\times}$.  
Hence,  the action for tensor perturbation is canonical and leads to  the 
standard MS equation:
	\begin{equation}
		\label{eq:MSeqtensor}
		\mu_k^{\prime \prime} + \left( c_T^2k^2 - \frac{a^{\prime \prime}}{a} 
\right) \mu_k = 0 ~,
	\end{equation}
	with  $\mu_k \equiv \frac{1}{2}a \gamma_k$   the MS variable.
	
	\subsection{Evolution of Perturbations in Ekpyrotic Contraction Phase}
	
	In the Ekpyrotic contracting phase  we have $|\phi| \gg 1$ and $\dot{\phi} 
\ll 1$, therefore $g \simeq 0$ and the higher order operators are suppressed. 
As analyzed in \cite{Cai:2012va}, we have
	\begin{equation}
		\label{eq:contractionz2c2}
		z_s^2 \simeq a^2/q ~,~ c_s^2 \simeq 1 ~,~ c_T^2 \simeq 1 ~.
	\end{equation}
	The validity of \eqref{eq:contractionz2c2} is also verified   numerically
 in \cite{Ilyas:2020qja}.
	
Using equation \eqref{eq:contractionaH} and 
\eqref{eq:contractionz2c2}  we obtain
	\begin{equation}
		\frac{z_s^{\prime \prime}}{z_s} = \frac{a^{\prime \prime}}{a} = 
\frac{q(2q-1)}{(1-q)^2 (\tau - \tau_-)^2} ~,
	\end{equation}
	hence the MS equations for scalar   \eqref{eq:MSeqscalar} and 
tensor perturbations \eqref{eq:MSeqtensor} become
	\begin{equation}
		\label{eq:MSeqscalarc}
		v_k^{\prime \prime} + \left[ k^2 - \frac{q(2q-1)}{(1-q)^2 (\tau - 
\tau_-)^2} \right] v_k = 0 ~,
	\end{equation}
	and
	\begin{equation}
		\mu_k^{\prime \prime} + \left[ k^2 - \frac{q(2q-1)}{(1-q)^2 (\tau - 
\tau_-)^2} \right] \mu_k = 0 ~.
	\end{equation}
	
	Let us first investigate  the scalar perturbations. The general 
analytical solution of \eqref{eq:MSeqscalarc} is
	\begin{equation}
		\label{eq:vkcontraction}
		v_k^c(\tau) = b_{s,1}(k) \sqrt{\tau_- - \tau} J_{\nu_c} [k(\tau_- - 
\tau)] + b_{s,2}(k) \sqrt{\tau_- - \tau} Y_{\nu_c} [k(\tau_- - \tau)] ~,
	\end{equation}
	with 
	\begin{equation}
		\nu_c = \frac{1-3q}{2(1-q)}=\frac{1}{2}-\frac{q}{1-q} ~.
	\end{equation}	
	Here, $J$ and $Y$ are Bessel functions, and we introduce the superscript 
``c'' in $v_k^c$ to denote the contraction phase. Recalling that, for small 
arguments, the Bessel function has the asymptotic behavior
	\begin{equation}
		J_a(x) \propto \left( \frac{x}{2} \right)^a ~,~ Y_a(x) \propto \left( 
\frac{2}{x} \right)^a ~, 0<x \ll \sqrt{a+1} ~,
	\end{equation}
we deduce that  the first part in the expression of $v_k^c$ 
\eqref{eq:vkcontraction} will be suppressed in the contraction phase. Hence, at 
the end of the Ekpyrotic contraction phase, the dominant contribution to $v_k$ 
comes from the second mode in \eqref{eq:vkcontraction}, and therefore we may 
set $b_{s,1} \simeq 0$ at that time.
	
Concerning  the tensor perturbations, a similar analysis provides the general 
analytical solution   as
	\begin{equation}
		\label{eq:nukcontraction}
		\mu_k^c(\tau) = b_{t,1}(k) \sqrt{\tau_- - \tau} J_{\nu_c} [k(\tau_- - 
\tau)] + b_{t,2}(k) \sqrt{\tau_- - \tau} Y_{\nu_c} [k(\tau_- - \tau)] ~,
	\end{equation}
	and thus its second mode will dominate at the end of Ekpyrotic contraction 
phase.
	
	\subsection{Evolution of Perturbations in Bounce Phase}
		\label{sec:ptbounce}
	
In the bounce phase, the parametrization \eqref{eq:bouncepara} gives (for 
the  details see \cite{Cai:2012va}):
	\begin{equation}
		\label{eq:za''}
		\frac{z^{\prime \prime}}{z} \simeq \Gamma + \frac{2}{T^2} + \left( 
2\Gamma^2 + \frac{6\Gamma}{T^2} + \frac{4}{T^4} \right) t^2 ~, \frac{a^{\prime 
\prime}}{a} \simeq \Gamma (\Gamma \tau^2 + 1) ~.
	\end{equation}
	Recalling that $\Gamma$ is of the order of $\mathcal{O}(10^{-4})$ and $T 
\simeq 0.5$,   we can approximately ignore the terms in \eqref{eq:za''} which
contain $\Gamma$. Thus, the MS equations simplify to  
	\begin{equation}
		\label{eq:MSeqSb}
		v_k^{\prime \prime} - \left( \frac{2}{T^2} + \frac{4\tau^2}{T^4} - c_s^2 
k^2 \right) v_k = 0 ~,
	\end{equation}
	\begin{equation}
		\label{eq:MSeqTb}
		\mu_k^{\prime \prime} + \left( c_T^2 k^2 - \Gamma \right) \mu_k = 0 ~.
	\end{equation}
	Note that in \eqref{eq:MSeqSb} we have used the approximation $\tau \simeq 
t$. The scale-dependent terms, $c_s^2k^2$ in equation \ref{eq:MSeqSb} and 
$c_T^2$ in equation \ref{eq:MSeqTb}, will be sub-dominant due to the smallness 
of the observable wavenumber $k$. Then, we can further neglect the $c_T^2k^2$ 
term in equation \ref{eq:MSeqTb}; however, we will maintain the $c_s^2k^2$ 
term in equation \ref{eq:MSeqSb} to examine whether it could explain the 
deviation of scalar spectra index $n_s$ from $1$, deduced from CMB data. 
	
	The final general solution for tensor perturbations is
	\begin{equation}
		\label{eq:nukb}
		\mu_k = b_{t,3} e^{\sqrt{\Gamma} \tau} + b_{t,4} e^{-\sqrt{\Gamma} \tau} 
~.
	\end{equation}
	However, $\sqrt{\Gamma} \tau_{B-}$ and $\sqrt{\Gamma} \tau_{B+}$ are of the
order of $10^{-2}$, therefore the exponential function in \eqref{eq:nukb} will 
be very close to $1$, which implies that the tensor perturbations will remain 
almost invariant during the bounce phase.
	
	The general solution for scalar perturbations is the parabolic cylinder 
function. We provide the detailed analysis in Appendix \ref{app:scalarbounce}, 
and here we   present the result. In particular, the dynamics of scalar 
perturbations can be approximated by the following semi-analytical solution:
	\begin{eqnarray}
&&		\label{eq:vkb}
		v_k^b(\tau) \simeq b_{s,3}(k)e^{\int \omega_k(\tau) d\tau} + 
b_{s,4}(k)e^{-\int \omega_k(\tau) d\tau} ~, \\
&&\omega_k^2(\tau) = 
\frac{4\tau^2}{T^4} + \frac{2}{T^2} - c_s^2(\tau) k^2 ~, 
	\end{eqnarray}
	where $b_{s,3}$ and $b_{s,4}$ can be determined by the matching condition. 
It would also be useful to define the quantity 
	\begin{equation}
		W_k(\tau) \equiv \int_{\tau_{B-}}^{\tau} \omega_k(x)dx ~,
	\end{equation}
	with
		\begin{equation}
		 \mathcal{F}_s(k) \equiv W_k(\tau_{B+}) ~, 
	\end{equation}
    and since $\mathcal{F}_s \simeq 6 \times 10^4$ is much larger than unity, 
the bounce phase will amplify the $b_{s,3}$ mode while it will suppress the 
$b_{s,4}$ mode. Hence, we can ignore the $b_{s,4}$ term in equation 
(\ref{eq:vkb}) at the end of the bounce phase. Besides, the $c_s^2(\tau)k^2$ 
term is sub-dominant in the expression of $\omega_k$, therefore the 
amplification factor $\mathcal{F}_s(k)$ is almost scale-independent, and the 
scalar spectra index will receive only a minor correction during the bounce 
phase.
	
	From the above argument  we can deduce that the tensor perturbations are 
almost invariant during the bounce phase, while the scalar perturbations get 
amplified by a nearly scale-independent factor $\mathcal{F}_s$. Hence, the 
existence of the bounce phase significantly suppresses the tensor-to-scalar 
ratio. As we will see in the following, this completely solves the   
problem of large tensor-to-scalar ratio of the usual matter bounce 
scenarios \cite{Cai:2008qw,Cai:2014xxa,Quintin:2015rta,Banerjee:2016hom}.

	\subsection{Evolution of Perturbations in Fast-roll Expansion Phase}
	
	After the bounce, the scalar field returns to the canonical form, and 
behaves like a perfect fluid with EoS parameter $w = 1$. As analyzed in 
\cite{Cai:2012va}, one has
	\begin{equation}
		z_s^2 \propto a^2 ~,~ c_s^2 \simeq 1 ~,~ c_T^2 \simeq 1 ~.
	\end{equation}
		Similarly, with the help of \eqref{eq:expansionah}, we obtain
	\begin{equation}
		\frac{z_s^{\prime \prime}}{z_s} = \frac{a^{\prime \prime}}{a} = 
-\frac{1}{4} \left( \frac{1}{\tau - \tau_+} \right)^2 ~,
	\end{equation}
	and thus the MS equations \eqref{eq:MSeqscalar} and \eqref{eq:MSeqtensor} 
become
	\begin{equation}
		v_k^{\prime \prime} + \left[ k^2 + \frac{1}{4} \frac{1}{(\tau - 
\tau_+)^2} \right] v_k = 0 ~,
	\end{equation}
	\begin{equation}
		\mu_k^{\prime \prime} + \left[ k^2 + \frac{1}{4} \frac{1}{(\tau - 
\tau_+)^2} \right] \mu_k = 0 ~.
	\end{equation}
		The general solution for the perturbations is then
	\begin{equation}
		\label{eq:vke}
		v_k^e(\tau) = b_{s,5} \sqrt{\tau - \tau_+} J_{\nu_e}[k(\tau - \tau_+)] + 
b_{s,6} \sqrt{\tau - \tau_+} Y_{\nu_e}[k(\tau - \tau_+)] ~,
	\end{equation}
	\begin{equation}
		\label{eq:muke}
		\mu_k^e(\tau) = b_{t,5} \sqrt{\tau - \tau_+} J_{\nu_e}[k(\tau - \tau_+)] 
+ b_{t,6} \sqrt{\tau - \tau_+} Y_{\nu_e}[k(\tau - \tau_+)] ~,
	\end{equation}
	with $\nu_e = 0$ ~.
	
	\subsection{Matching Conditions and   Power Spectra}

	After obtaining general solutions to the perturbation equations in the 
various phases of the cosmological evolution, we now determine how the 
solutions should be matched at the transition surfaces $\tau = \tau_{B-}$ and 
$\tau = \tau_{B+}$. It was argued in \cite{Cai:2012va} that the matching 
condition for scalar perturbations, deduced from 
\cite{Hwang:1991an,Deruelle:1995kd} (see also related discussions in \cite{Creminelli:2012my,Hinterbichler:2012yn, Padilla:2012ze,Nishi:2014bsa,Aviles:2019xae}), is the continuity of $v$ and $v^{\prime}$ 
at the transition surface. The same argument can be applied to tensor 
perturbations (as we   show in Appendix \ref{app:tensormatch}), and $\mu$, 
$\mu^{\prime}$ are also continuous across the transition surface.
	
	For scalar perturbations, the coefficients of   \eqref{eq:vke} are 
	\begin{equation}   
		b_{s,5} = \frac{\mathcal{F}_{s} 
\Gamma\left(\nu_{c}\right)\left(1-2\nu_c\right)^{\frac{1}{2}-\nu_{c}}}{4^{1-\nu_
{c}}\pi}\left(\frac{-\mathcal{H}_{B-}}{k}\right)^{\nu_{c}} 
\frac{w_{k+}}{\sqrt{-\mathcal{H}_{B-} \mathcal{H}_{B+}}}\ln 
\frac{ke^{\gamma_E}}{4 \mathcal{H}_{B+}} b_{s, 2} ~,
	\end{equation}
	\begin{equation}   
		b_{s,6} = -\frac{\mathcal{F}_s \Gamma(\nu_c) 
(1-2\nu_c)^{\frac{1}{2}-\nu_c}}{2^{3-2\nu_c}} \left( \frac{-\mathcal{H_{B-}}}{k} 
\right)^{\nu_c} \frac{\omega_{k+}}{\sqrt{-\mathcal{H_{B-}} \mathcal{H_{B+}} }} 
b_{s,2} ~,
	\end{equation}
	where the involved calculation details are presented 
  in Appendix \ref{app:scalarmatch}. Here  $\Gamma(x)$ is 
the gamma function and $\gamma_E \simeq 0.58$ is the Euler-Masheroni constant. 
The expression of $v_k^e$ as a function of the conformal Hubble parameter 
$\mathcal{H}$ is then
	\begin{equation}
		\label{eq:vkefinal}
		v_k^e(\mathcal{H}) = \frac{\mathcal{F}_{s} 
\Gamma\left(\nu_c\right)\left(1-2\nu_c\right)^{\frac{1}{2}-\nu_c}}{4^{1-\nu_c} 
\pi \sqrt{ 2\mathcal{H}}} \left( \frac{-\mathcal{H_{B-}}}{k} \right)^{\nu_c} 
\left( \frac{\omega_{k+} \ln \frac{\mathcal{H}}{\mathcal{H}_{B+} }} 
{\sqrt{-\mathcal{H_{B-}} \mathcal{H_{B+}} }}  - 
2\sqrt{\frac{\mathcal{H_{B+}}}{-\mathcal{H_{B-}}}} \right) b_{s,2} ~,
	\end{equation}
	where we have used \eqref{eq:tau+}.
	
	At the beginning of the fast-roll expansion phase, the wavenumber $k$ of 
observational interest lies outside the horizon since $k < \mathcal{H}_{B+}$. 
We will consider that the scalar perturbation is always outside the horizon 
during this phase. Standard cosmology implies that   scalar 
perturbations should cross the horizon after the beginning of Big Bang 
Nucleosynthesis (BBN), hence the wavenumber observed by Large Scale Structure 
(LSS) must stay outside the horizon \cite{Saridakis:2021qxb}. We can also see 
this by a simple estimation. If a certain mode $k$ crosses the horizon during 
the fast-roll 
expansion phase, then from $k = \mathcal{H}_{\star}$ and the formulae of 
subsection \ref{sec:bgdynamicse} we can acquire the corresponding energy density 
of the Universe $\rho_{\star}$ as
	\begin{equation}
	\label{eq:rhostar}
		\rho_{\star} = 3H_{\star}^2 = 3 \left[ \frac{(\tau_{B+} - 
\tau_+)^{\frac{3}{2}}}{3(t_{B+} - t_+)} \times 
(2\mathcal{H}_{\star})^{\frac{3}{2}} \right]^2 = \frac{3k^3}{\mathcal{H}_{B+}} 
~.
	\end{equation}
	The wavenumber $k$ of observational interest has an upper limit $k < 
10^{-10}$ in SI units. With the help of \eqref{eq:rhostar}, the corresponding energy density is $\rho_{\star} \simeq \mathcal{O}(10^{-135})$ in Planck units,
which is far below the energy scale of BBN. Hence, it is no possible for the 
scalar perturbations to cross the horizon during the fast-roll expansion phase.

Denoting the conformal Hubble parameter at the end of fast-roll expansion 
phase as $\mathcal{H}_r$, the definition of scalar power spectra
	\begin{equation}
	    P_{\zeta} \equiv \frac{k^3}{2\pi^2}|\zeta_k|^2 = \frac{k^3}{6\pi^2} | 
\frac{v_k^e}{a} | ^2 ~,
	\end{equation}
	leads to (we use the approximation $a_{B+} \simeq 1$ and $a_{B-} 
\simeq 1$)
	\begin{equation}
		\label{eq:Ps}
		P_{\zeta} = \frac{\mathcal{F}_s^2 \Gamma^2(\nu_c) (1-2\nu_c)^{1-2\nu_c} 
}{3\pi^4 2^{6-4\nu_c}} \frac{(-\mathcal{H_{B-}})^{2\nu_c - 1}}{k^{2\nu_c-3}}  
\left( \frac{\omega_{k+}}{\mathcal{H_{B+}}} \ln 
\frac{\mathcal{H}}{\mathcal{H_{B+}}} - 2 \right)^2 |b_{s,2}|^2 ~.
	\end{equation}
	Therefore the  spectra index becomes	
	\begin{equation}
		\label{eq:ns}
		n_s - 1 \equiv \frac{d\ln P_{\zeta}}{d\ln k} \simeq 3-2\nu_c + 
\frac{d\ln |b_{s,2}|^2}{d\ln k} ~,
	\end{equation}
 where we have omitted  the k-dependence of $\omega_{k+}$ and 
$\mathcal{F}_s(k)$ sine it is negligible. Equations \eqref{eq:Ps} and 
\eqref{eq:ns} are the main results for the scalar perturbations.
	
Let us follow the same procedure for the tensor perturbations. The 
coefficients of expression \eqref{eq:muke} are
	\begin{equation}
		\frac{b_{t,6}}{b_{t,2}} = \frac{\left(1-2 
\nu_{c}\right)^{\frac{1}{2}-\nu_{c}} \Gamma\left(\nu_{c}\right)}{4^{1-\nu_{c}}} 
\sqrt{\frac{\mathcal{H}_{B+}}{-\mathcal{H}_{B-}}}\left(1+\frac{-\mathcal{H}_{B-}
}{\mathcal{H}_{B+}}\right)\left(\frac{-\mathcal{H}_{B-}}{k}\right)^{\nu_{c}}  ~,
	\end{equation}
	\begin{equation}
		\frac{b_{t,5}}{b_{t,2}} = \frac{-\left(1-2 
\nu_{c}\right)^{\frac{1}{2}-\nu_{c}} \Gamma\left(\nu_{c}\right)}{2^{1-2 \nu_{c}} 
\pi} 
\sqrt{\frac{\mathcal{H}_{B+}}{-\mathcal{H}_{B-}}}\left[2+\left(1+\frac{-\mathcal
{H}_{B-}}{\mathcal{H}_{B+}}\right)\ln \frac{ke^{\gamma_E}}{4 \mathcal{H}_{B+}} 
\right]\left(\frac{-\mathcal{H}_{B-}}{k}\right)^{\nu_{c}} ~,
	\end{equation}
the detailed calculations are presented in Appendix 
\ref{app:tensormatch}. Additionally, relation \eqref{eq:muke} becomes:
	\begin{equation}
		\mu_k^e(\mathcal{H}) = \frac{\Gamma(\nu_c) (1-2\nu_c)^{\frac{1}{2} 
-\nu_c}} {2^{1-2\nu_c} \pi \sqrt{2\mathcal{H}}} 
\sqrt{\frac{\mathcal{H_{B+}}}{-\mathcal{H_{B-}}}} \left( 
\frac{-\mathcal{H_{B-}}}{k} \right)^{\nu_c}  \left[ \left( 1 + 
\frac{-\mathcal{H_{B-}}}{\mathcal{H_{B+}}} \right) \ln 
\frac{\mathcal{H}_{B+}}{\mathcal{H}} - 2 \right] b_{t,2} ~.
	\end{equation}
	Hence, the final tensor power spectrum, defined by $P_t = 2 \times 
\frac{k^3}{2\pi^2} | \frac{\mu_k^e}{a/2} | ^2$, is
	\begin{equation}
	\label{eq:Pt}
		P_t = \frac{\Gamma^2(\nu_c) (1-2\nu_c)^{1-2\nu_c}}{2^{1-4\nu_c} \pi^4 } 
\frac{(-\mathcal{H_{B-}})^{2\nu_c-1}}{k^{2\nu_c-3}} \left[ \left( 1 + 
\frac{-\mathcal{H_{B-}}}{\mathcal{H_{B+}}} \right) \ln 
\frac{\mathcal{H}_{B+}}{\mathcal{H}_r} - 2 \right]^2 |b_{t,2}|^2 ~,
	\end{equation}
therefore the tensor spectral index $n_t$ and the tensor-to-scalar 
ratio $r$ become
	\begin{equation}
    \label{eq:nt}	
		n_t = \frac{d\ln P_t}{d\ln k} = 3-2\nu_c + \frac{d\ln |b_{t,2}|^2}{d\ln 
k} ~,
	\end{equation}
	\begin{equation}
	\label{eq:r}
		r \equiv \frac{P_t}{P_{\zeta}} = \frac{96}{\mathcal{F}_s^2} \frac 
{\left[ \left( 1 + \frac{-\mathcal{H_{B-}}} {\mathcal{H_{B+}}} \right) \ln 
\frac{\mathcal{H}_{B+}} {\mathcal{H}_r} - 2 \right]^2} {\left( 
\frac{\omega_{k+}}{\mathcal{H_{B+}}} \ln \frac{\mathcal{H}_{B+}} {\mathcal{H}_r} 
+ 2 \right)^2} \left|\frac{b_{t,2}}{b_{s,2}}\right|^2~.
	\end{equation}
	Equations \eqref{eq:Pt}, \eqref{eq:nt} and \eqref{eq:r} are the main 
results for the tensor perturbations. 

	\section{Initial Conditions and Power Spectra}
	\label{sec:initial}
	
In the previous sections we have presented the core of calculations for the 
scalar and tensor perturbations. A key issue that needs to be incorporated 
is how to choose proper initial conditions for these 
perturbations. In the following subsections we consider three sets of initial 
conditions, and determine the corresponding expressions for the 
power spectra.
	
	\subsection{Perturbations from Ekpyrotic Vacuum Fluctuations}
	\label{sec:spectrae}
	
	Firstly, we consider the cosmological perturbations arising from the 
quantum vacuum fluctuations at the very beginning of the universe. In the 
Ekpyrotic contraction phase the action for $v_k$ is of the standard harmonic 
oscillator form, thus at far past we have
	\begin{equation}
		v_k(\tau) \to \frac{e^{-ik(\tau - \tau_-)}}{\sqrt{2k}} ~,~ \tau \to 
-\infty ~.
	\end{equation}
	Applying the small-argument approximation of the Bessel function, 
expression \eqref{eq:vkcontraction} becomes
	\begin{equation}
		v_k(\tau) \to b_{s,1} \sqrt{\frac{2}{\pi k}} \cos \left[ k(\tau_- - 
\tau) - \frac{\pi}{2}\nu_c - \frac{\pi}{4} \right] + b_{s,2} \sqrt{\frac{2}{\pi 
k}} \sin \left[ k(\tau_- - \tau) - \frac{\pi}{2}\nu_c - \frac{\pi}{4} \right] ~,
	\end{equation}
	which gives
	\begin{equation}
		\label{eq:bs2evacuum}
		b_{s,2} = \frac{\sqrt{\pi}}{2} e^{i\left( \frac{\pi}{2}\nu_c + 
\frac{3\pi}{4} \right)} ~ ~ \mbox{or} ~ |b_{s,2}| = \frac{\sqrt{\pi}}{2} ~.
	\end{equation}
	Substituting   \eqref{eq:bs2evacuum} into the expression of scalar 
power spectrum \eqref{eq:Ps}, we obtain
	\begin{equation}
		\label{eq:Pse}
		P_{\zeta} = \frac{\mathcal{F}_s^2 \Gamma^2(\nu_c) (1-2\nu_c)^{1-2\nu_c} 
}{3\pi^3 2^{8-4\nu_c}} \frac{(-\mathcal{H_{B-}})^{2\nu_c - 1}}{k^{2\nu_c-3}}  
\left( \frac{\omega_{k+}}{\mathcal{H_{B+}}} \ln 
\frac{\mathcal{H_{B+}}}{\mathcal{H}_r} + 2 \right)^2 ~.
	\end{equation}
	Hence, we deduce that the spectra index is $n_s-1 = 3-2\nu_c$. However, 
the constraint $0<q<1/3$ gives $0<\nu_c<\frac{1}{2}$ and thus $2<n_s-1<3$. 
Therefore, the scalar power spectrum generated by   quantum fluctuations 
during the Ekpyrotic phase is always blue \footnote{The blue spectrum is a 
generic feature for artificially smoothed-out four-dimensional basic models of 
Ekpyrotic cosmology \cite{Lyth:2001pf, Brandenberger:2001bs,Tsujikawa:2002qc}. 
Nevertheless, the scale invariance can be acquired in Ekpyrotic scenario by 
applying string theory considerations 
\cite{Tolley:2003nx,Notari:2002yc,Finelli:2002we, 
Creminelli:2007aq,Lehners:2007ac,Buchbinder:2007ad, Battefeld:2005wv}.}, which 
is not favored by the current observations.
	
The same procedure can be applied to the tensor perturbations that leads to  
$|b_{t,2}| = \frac{\sqrt{\pi}}{2}$, and 
	\begin{equation}
		P_t = \frac{\Gamma^2(\nu_c) (1-2\nu_c)^{1-2\nu_c} }{\pi^3 2^{3-4\nu_c}} 
\frac{(-\mathcal{H_{B-}})^{2\nu_c-1}}{k^{2\nu_c-3}} \left[ \left( 1 + 
\frac{-\mathcal{H_{B-}}}{\mathcal{H_{B+}}} \right) \ln 
\frac{\mathcal{H}_{B+}}{\mathcal{H}_r} - 2 \right]^2 ~.
	\end{equation}
	Hence, we deduce that 
	\begin{equation}
		n_t = 3-2\nu_c ~,\ \ \ \ 2<n_t<3 ~.
	\end{equation}
	Similarly to the scalar case, the tensor power spectrum is always blue. 
Hence, we conclude that if the cosmological perturbations are generated by 
Ekpyrotic vacuum fluctuations, then the resulting scalar and tensor spectra are 
always blue, and thus in principle this case is ruled out due to observations.

	\subsection{Perturbations from Matter Vacuum Fluctuations}

	In a realistic bouncing universe  there should be matter and radiation 
sectors at the end of the bounce phase, in order for the model to 
successfully transit into standard cosmology. If there is no matter and 
radiation generated during the bounce phase, as it is the case in our scenario, 
then at the distant past there should be a phase during which matter and 
radiation would have been dominant. The perturbations then arise from the initial quantum fluctuations of the matter field, since the EoS parameter of matter is 
smaller than that of radiation. Moreover, it is well known that 
the spectra generated in the matter bounce 
scenario are scale invariant as long as the perturbation modes exit the Hubble 
radius during the matter contraction phase \cite{Cai:2008qw, Wands:1998yp, Durrer:2002jn}.  Thus, it 
would be important to determine how the 
cosmological perturbations evolve in the situation where a matter contraction 
phase takes place before the Ekpyrotic contraction phase in our model.
	
	In principle, in order to fully address the issue one needs to include the 
effect of matter and radiation on the dynamics of background and perturbations, 
as it was done in \cite{Cai:2013kja}. Here we would like to consider particular 
simplifications. Firstly, we focus on   adiabatic perturbations since the 
possible generated entropy perturbations are highly dependent on the matter 
field. Moreover, we assume that the dynamics of matter contraction and 
Ekpyrotic contraction phases are solely determined by the matter and 
DHOST fields, namely the effects of other fields on these phases are 
negligible. Under these assumptions, we can approximately elaborate the power 
spectra, which may provide some insights on the complete analysis. 
	
	With the above assumptions, the dynamics of perturbations are fully 
investigated in \cite{Cai:2012va}, and the forms of $b_{s,2}$ and $b_{t,2}$ are
	\begin{equation}
		b_{s,2} = b_{t,2} \simeq \frac{-\pi k^{\nu_c - \frac{3}{2}}(\tau_m - 
\tau_-)^{\nu_c - \frac{1}{2}} \mathcal{H}_m}{2^{\nu_c + \frac{3}{2}} 
\Gamma(\nu_c) } ~,
	\end{equation}
	where $\tau_m$ and $\mathcal{H}_m$ respectively are the conformal time and 
Hubble parameter at the end of matter contraction phase. The power 
spectra of scalar and tensor perturbation are then
	\begin{equation}
		\label{eq:Psm}
		P_{\zeta} = \frac{\mathcal{F}_s^2 \mathcal{H}_m^2}{3\pi^2 2^{9-2\nu_c}}   \left( \frac{\omega_{k+}}{\mathcal{H_{B+}}} \ln \frac{\mathcal{H_{B+}}}{\mathcal{H}_r} + 2 \right)^2 \left( -\mathcal{H_{B-}} \frac{\tau_m - \tau_-}{1-2\nu_c} \right)^{2\nu_c - 1} ~, 
	\end{equation}
	and
	\begin{equation}
	\label{eq:Ptm}
		P_t = \frac{\left[ \left( 1 + \frac{-\mathcal{H_{B-}}} 
{\mathcal{H_{B+}}} \right) \ln \frac{\mathcal{H}_{B+}} {\mathcal{H}_r} - 2 
\right]^2}{\pi^2 2^{4-2\nu_c}} \left( -\mathcal{H_{B-}} \frac{\tau_m - 
\tau_-}{1-2\nu_c} \right)^{2\nu_c - 1} \mathcal{H}_m^2 ~.
	\end{equation}
	
	From expressions \eqref{eq:Psm} and \eqref{eq:Ptm}  we see that the scalar 
spectra are scale invariant, which is in agreement with   observations.   The 
tensor-to-scalar ratio is given as
	\begin{equation}
		r \equiv \frac{P_t}{P_{\zeta}} =  \frac{96}{\mathcal{F}_s^2} \frac 
{\left[ \left( 1 + \frac{-\mathcal{H_{B-}}} {\mathcal{H_{B+}}} \right) \ln 
\frac{\mathcal{H}_{B+}} {\mathcal{H}_r} - 2 \right]^2} {\left( 
\frac{\omega_{k+}}{\mathcal{H_{B+}}} \ln \frac{\mathcal{H_{B+}}}{\mathcal{H}_r} 
+ 2 \right)^2} ~,
	\end{equation}
	and the spectral index is $n_s = 1$. Comparing to the observational data 
$n_s = 0.9649 \pm 0.0042$ \cite{Aghanim:2018eyx}, we need a complete 
investigation on the matter bounce scenario to generate the small tilt $n_s - 1 
\simeq -0.04$.
	
	\subsection{Perturbations from Thermal Fluctuations}

	The primordial scalar perturbations could arise from thermal fluctuations of a thermal ensemble of point particles, i.e. relativistic particles with EoS 
parameter $w_r = 1/3$. Note that the thermal gas cannot be the source of tensor 
perturbations, thus we still have $|b_{t,2}|^2 = \pi/4$, and hence the tensor 
spectrum becomes
	\begin{equation}
	\label{eq:Ptt}
		P_t = \frac{\Gamma^2(\nu_c) (1-2\nu_c)^{1-2\nu_c}  }{\pi^3 2^{3-4\nu_c}} 
\frac{(-\mathcal{H_{B-}})^{2\nu_c-1}}{k^{2\nu_c-3}} \left[ \left( 1 + 
\frac{-\mathcal{H_{B-}}}{\mathcal{H_{B+}}} \right) \ln 
\frac{\mathcal{H}_{B+}}{\mathcal{H}_r} - 2 \right]^2 ~.
	\end{equation}
	
	Thermal fluctuations in bouncing cosmologies have been well investigated in 
\cite{Cai:2009rd}. For a gas of point particles with an EoS parameter $w_r = 
1/3$, its energy density and temperature evolve  as
	\begin{equation}
		\rho_r = \rho_{r,B} a^{-4} ~,\ \ \  T = T_B a^{-1} ~,
	\end{equation}
	where $\rho_{r,B}$ and $T_B$ are the energy density and temperature at the bounce point respectively, and here we treat them as model parameters. The heat capacity is
	\begin{equation}
		C_V(R) = R^3 \frac{\partial \rho_r}{\partial T} = R^3 \left( \frac{\partial T}{\partial a} \right)^{-1} \left( \frac{\partial \rho_r}{\partial a} \right) = \frac{4\rho_{r,B}}{T_B^4}  R^3 T^3 ~,
	\end{equation}
	where $R$ is the linear size of the thermal gas  which is approximated by 
the Hubble radius $R \sim 1/H$. The correlation function of the energy density 
is given by
	\begin{equation}
		\langle \delta \rho^2 \rangle_{|R(k)} = C_V(R) \frac{T^2}{R^6} = 
\frac{4\rho_{r,B}}{T_B^4} T^5H^3  = 4\rho_{r,B} T_B H^3a^{-5} ~. 
	\end{equation}
 The MS variable at the horizon crossing point  $v_k(\tau_H)$ can be related to 
the density fluctuation by \cite{Cai:2012va}
	\begin{equation}
		|v_k(\tau_H)| = \frac{a^2 H \delta \rho_k}{2k \sqrt{q} \dot{H}} \simeq \frac{a^2 H k^{-\frac{3}{2}}}{2k \sqrt{q} \dot{H}} \sqrt{ \langle \delta \rho^2 \rangle } = \frac{\sqrt{\rho_{r,B} T_B}}{\epsilon_H \sqrt{q}}  \frac{1}{k^2}\left(\frac{k}{-\mathcal{H}_{B-}} \right)^{\frac{1}{2}-\nu_c} ~,
	\end{equation}
    where we have used the fact that the effective ``slow-roll parameter'' 
$\epsilon_H \equiv -\dot{H}/H^2$ is approximately a constant.
	
	Finally, with the background dynamics in Ekpyrotic contraction phase 
\eqref{eq:contractionaH} we acquire the expression for the initial conditions as
	\begin{equation}
		|b_{s,2}| = \frac{\pi k^{\frac{1}{2}}}{2^{\nu_c} \Gamma(\nu_c)}  |v_k(\tau_H)| = \frac{\pi \sqrt{\rho_{r,B} T_B}}{\epsilon_H \sqrt{q} 2^{\nu_c} \Gamma(\nu_c)} \left( -\mathcal{H}_{B-} \right)^{\frac{1}{2} - \nu_c} k^{-1-\nu_c} ~,
	\end{equation}
	and the scalar power spectrum is
	\begin{equation}
	\label{eq:Pst}
		P_{\zeta} = \frac{\mathcal{F}_s^2 \rho_{r,B}T_B (1-2\nu_c)^{1-2\nu_c}} 
{3\pi^2 \epsilon_H^2 q 2^{6-2\nu_c}}  \left(\frac{\omega_{k+}}{\mathcal{H_{B+}}} 
\ln \frac{\mathcal{H}_{B+}}{\mathcal{H}_r} + 2 \right)^2 k^{1-4\nu_c} ~.
	\end{equation}
Finally, the tensor-to-scalar ratio is
	\begin{equation}
	\label{eq:rthermalcase}
		r = \frac{3q\Gamma^2(\nu_c) \epsilon_H^2 2^{3+2\nu_c}  k^{2+2\nu_c}}{\pi \mathcal{F}_s^2 (-\mathcal{H_{B-}})^{1-2\nu_c} \rho_{r,B} T_B} \frac{\left[ \left( 1 + \frac{-\mathcal{H_{B-}}}{\mathcal{H_{B+}}} \right) \ln \frac{\mathcal{H}_{B+}}{\mathcal{H}_r} - 2 \right]^2} {\left(\frac{\omega_{k+}}{\mathcal{H_{B+}}}  \ln \frac{\mathcal{H}_{B+}}{\mathcal{H}_r} + 2 \right)^2} ~.
	\end{equation}
	
    From expression \eqref{eq:Pst} we immediately deduce that   spectral index 
is $n_s - 1 = 1 - 4\nu_c$. To confront with observational constraints, we have
    \begin{equation}
    	1-4\nu_c = -0.04 ~\to~ \nu_c = 0.26 ~,~ q = \frac{6}{31} ~,
    	\label{parameterresults}
    \end{equation}
    which explains the values given in \eqref{eq:parameter}.
    
    Lastly, the corresponding tensor spectral index is $n_t = 3-2\nu_c = 
2.48$, and thus  the tensor spectrum is always blue, which is different from 
the matter bounce case.

	\subsection{Summary of Results}
	We have analyzed the power spectra with three sets of initial conditions. 
The cosmological perturbations generated by the vacuum quantum fluctuation in 
the Ekpyrotic contraction phase will always result in a blue scalar spectrum 
with spectra index $n_s = 2/(1-q) > 3$, which is ruled out by  
observations. The other two cases can provide a nearly scale-invariant scalar 
power spectrum, hence they satisfy the basic requirement to be 
consistent with observations. For these two case, we summarize the main results 
below.
	\begin{itemize}
		\item Matter Vacuum Fluctuations
		\begin{itemize}
			\item $n_s \simeq 0$ Condition: Exit horizon at matter contraction 
phase
			\item Amplitude: $\frac{\mathcal{F}_s^2 \mathcal{H}_m^2}{3\pi^2 2^{9-2\nu_c}} \left( \frac{\omega_{k+}}{\mathcal{H_{B+}}} \ln \frac{\mathcal{H_{B+}}}{\mathcal{H}_r} + 2 \right)^2 \left( -\mathcal{H_{B-}} \frac{\tau_m - \tau_-}{1-2\nu_c} \right)^{2\nu_c - 1}$
			\item Tensor spectrum: Scale Invariant
			\item r: $\frac{96}{\mathcal{F}_s^2} \frac {\left[ \left( 1 + \frac{-\mathcal{H_{B-}}} {\mathcal{H_{B+}}} \right) \ln \frac{\mathcal{H}_{B+}} {\mathcal{H}_r} - 2 \right]^2} {\left( \frac{\omega_{k+}}{\mathcal{H_{B+}}} \ln \frac{\mathcal{H_{B+}}}{\mathcal{H}_r} + 2 \right)^2}$
			\item $n_s - 1$: 0
		\end{itemize}

		\item Thermal Fluctuations
		\begin{itemize}
			\item $n_s \simeq 0$ Condition: $q \simeq \frac{1}{5}$. When $q = 
\frac{6}{31}$ and $\nu_c = 0.26$, $n_s - 1 = -0.04$ is consistent with 
observations.
			\item Amplitude: $\frac{\mathcal{F}_s^2 \rho_{r,B}T_B (1-2\nu_c)^{1-2\nu_c}}{3\pi^2 \epsilon_H^2 q 2^{6-2\nu_c}}  \left( \frac{\omega_{k+}}{\mathcal{H_{B+}}}\ln \frac{\mathcal{H}_{B+}}{\mathcal{H}_r} + 2 \right)^2 k^{1-4\nu_c}$
			\item Tensor spectrum: Blue with $n_t = 2.48$
			\item r: $\frac{3q\Gamma^2(0.26) \epsilon_H^2 2^{3.52} k_{\max}^{2.52}}{\pi  \mathcal{F}_s^2 (-\mathcal{H_{B-}})^{0.48} \rho_{r,B} T_B} \frac{\left[ \left( 1 + \frac{-\mathcal{H_{B-}}}{\mathcal{H_{B+}}} \right) \ln \frac{\mathcal{H}_{B+}}{\mathcal{H}_r} - 2 \right]^2} {\left( \frac{\omega_{k+}}{\mathcal{H_{B+}}}\ln \frac{\mathcal{H}_{B+}}{\mathcal{H}_r} + 2 \right)^2}$
			\item $n_s - 1$: -0.04
		\end{itemize}
	\end{itemize}
	
	Finally, for readers' convenience  we   briefly summarize the 
definition of the physical quantities that appear in the above results.
	\begin{itemize}
		\item $q$:  the parameter describing the background dynamics of 
Ekpyrotic contraction phase. Is is related to the EoS parameter at Ekpyrotic 
phase $w_c$ by $w_c = -1 + \frac{2}{3q}$. $q$ ranges from $0$ to $1/3$, 
therefore the auxiliary parameter $\nu_c \equiv \frac{1-3q}{2(1-q)}$ ranges 
from $0$ to $\frac{1}{2}$.

		\item $\omega_k$, $\mathcal{F}_s$: the function $\omega_k(\tau)$, 
defined by $\omega_k^2(\tau) \equiv \frac{4\tau^2}{T^2} + \frac{2}{T^2} - 
c_s^2(\tau)k^2$, determines the dynamics of scalar perturbations $v_k$ during 
the bounce  phase. During this phase the scalar perturbations will get 
amplified by a factor $\mathcal{F}_s \equiv \exp \left( 
\int_{\tau_{B-}}^{\tau_{B+}} \omega_k(\tau) d\tau \right)$. Furthermore, 
$\omega_{k+} \equiv \omega_k(\tau_{B+})$. Although the quantities $\omega_k$ 
and $\mathcal{F}_s$ depend on $k$, their contributions to spectra index are 
negligible.

		\item $\mathcal{H}_r$: the conformal Hubble parameter at the beginning 
of radiation dominated phase.
		\item $\tau_m$, $\mathcal{H}_m$: these are the conformal time and 
Hubble parameter at the end of matter contraction, respectively.
		\item $\rho_{r,B}$, $T_B$:  these are the density and temperature of 
the thermal gas at the bounce point $\tau = 0$, respectively.
		\item $\epsilon_H$: the ``slow roll parameter'' $\epsilon_H \equiv 
-\dot{H}/H^2$ at the Ekpyrotic contraction phase, which is almost   constant.
	\end{itemize}
	
	Lastly, we summarize the approximations used throughout the analytical 
calculations. Our results will be valid for any single bounce models, as long 
as these approximations hold.

	\begin{itemize}
	    \item During the Ekpyrotic contraction and fast-roll expansion phases, 
the Lagrangian of the scalar field $\phi$ is approximately of the canonical 
form: $\mathcal{L} \to   \partial_{\mu}\phi \partial^{\mu} \phi/2$. Hence, the 
term $z_s^{\prime \prime}/z_s$, which appears in the MS equation for scalar 
perturbations, acquires the  simple expression $z_s^{\prime \prime}/z_s \simeq 
a^{\prime \prime}/a$.
	    \item The effective scalar potential $V(\phi)$ during the Ekpyrotic 
contraction  phase admits a specific attractor solution, leading to a 
nearly constant EoS parameter during that phase. 

	    \item The bounce phase is assumed to be short, which allows for the 
linear approximation $H = \Gamma t$. Moreover, the fact that the conformal time 
at the beginning/end of the bounce phase is small enables us to apply the 
small-argument approximation of the Bessel functions.

	    \item The scale-dependent terms, $c_s^2 k^2$ and $c_T^2 k^2$, are 
negligible in MS equations due to the small values of the observed wavenumber 
$k$.
	\end{itemize}

    \subsection{Comparison with   Observations}
    
    In this final subsection  we   consider the most efficient case, where  
cosmological  perturbations are  generated  by thermal fluctuations, and we 
present three different examples  compatible with cosmological observations.
    
    Firstly, in order to calculate the conformal wavenumber $k$  corresponding 
to the physical large-scale structure, we need to know the present scale factor 
$a_0$. With the useful formula 
  $a\propto H^{-\frac{1}{\epsilon}}$,
 we obtain
\begin{align}
a_0=&\frac{a_0}{a_{eq}}\frac{a_{eq}}{a_r}\frac{a_r}{a_{B+}}\frac{a_{B+}}{a_B} 
a_B\nonumber\\    
=&(1+z_{eq})\bigg(\frac{H_r}{H_{eq}}\bigg)^{\frac{1}{\epsilon_r}}\bigg(\frac{H_{B+}}{H_r}\bigg)^{ 
\frac{1}{\epsilon_{fr}}}, 
\end{align} 
where the subscript ``$eq$'' denotes the radiation-matter equality epoch 
(e.g. $H_{eq}$ is the Hubble parameter at the radiation-matter equality 
epoch).

We can continue writing $H_{eq}$ as $ H_{eq}=H_0 (1+z_{eq})^{\epsilon_{m}}$,  
and then we have
\begin{equation}
a_0=(1+z_{eq})^{1-\frac{\epsilon_m}{\epsilon_r}}\left(\frac{H_r}{H_{0}}\right)^{\frac{1}{\epsilon_r}}\left(\frac{H_{B+}}{H_r}\right)^{\frac{1}{\epsilon_{fr}}},
\end{equation}
where $\epsilon_m$, $\epsilon_r$ and $\epsilon_{fr}$ are the ``slow-roll'' parameters for matter, radiation and fast-roll phase, respectively. Recalling that $a_{B+}\simeq a_B=1$  and  that
$\epsilon_{m}=3/2,\epsilon_{r}=2,\epsilon_{fr}=3$, we have
\begin{equation}
    a_0=(1+z_{eq})^{\frac{1}{4}}\left(\frac{H_r}{H_{0}}\right)^{\frac{1}{2}}\left(\frac{H_{B+}}{H_r}\right)^{\frac{1}{3}}.
\end{equation}
Note that the relation between conformal wavenumber and physical wavenumber is 
 $k=K a_0$.

\begin{table}

    \begin{minipage}{0.2\linewidth}
    	\renewcommand{\arraystretch}{1.5}
    	\centering
    	\setlength{\tabcolsep}{0.8mm}
    \begin{tabular}{|c|c|c|c|}
	\hline
	\multicolumn{4}{|c|}{$p = 0.01 ~,~ g_0 = 1.1$} \\
	\hline
	$q$ & 0.193 & 0.194 & 0.195 \\
	\hline
	$n_s$ & 0.9566 & 0.9628 & 0.9689 \\
	\hline
	$10^{62}r$ & 4.09 & 4.00 & 4.40 \\
	\hline
    \end{tabular}
    \end{minipage}
    \hfill
    \begin{minipage}{0.24\linewidth}
    	\renewcommand{\arraystretch}{1.5}
    	\centering
    	\setlength{\tabcolsep}{0.8mm}
    \begin{tabular}{|c|c|c|c|}
    	\hline
    	\multicolumn{4}{|c|}{$p = 0.01 ~,~ g_0 = 1.2$} \\
    	\hline
    	$q$ & 0.193 & 0.194 & 0.195 \\
    	\hline
    	$n_s$ & ~0.9566~ & ~0.9628~ & ~0.9689~ \\
    	\hline
    	$10^{62}r$ & 2.93 & 3.03 & 3.50 \\
    	\hline
    \end{tabular}
    \end{minipage} 
    \hfill
    \begin{minipage}{0.33\linewidth}
    	\renewcommand{\arraystretch}{1.5}
    	\centering
    	\setlength{\tabcolsep}{0.8mm}
    	\begin{tabular}{|c|c|c|c|}
    		\hline
    		\multicolumn{4}{|c|}{$p = 0.01 ~,~ g_0 = 1.3$} \\
    		\hline
    		$q$ & 0.193 & 0.194 & 0.195 \\
    		\hline
    		$n_s$ & 0.9566 & 0.9628 & 0.9689 \\
    		\hline
    		$10^{62}r$ & 2.44 & 3.26 & 3.28 \\
    		\hline
    	\end{tabular}
    \end{minipage}

    \begin{minipage}{0.2\linewidth}
    	\renewcommand{\arraystretch}{1.5}
    	\centering
    	\setlength{\tabcolsep}{0.8mm}
    	\begin{tabular}{|c|c|c|c|}
    		\hline
    		\multicolumn{4}{|c|}{$p = 0.02 ~,~ g_0 = 1.1$} \\
    		\hline
    		$q$ & 0.193 & 0.194 & 0.195 \\
    		\hline
    		$n_s$ & 0.9566 & 0.9628 & 0.9689 \\
    		\hline
    		$10^{62}r$ & 3.31 & 3.78 & 3.56 \\
    		\hline
    	\end{tabular}
    \end{minipage}
    \hfill
    \begin{minipage}{0.24\linewidth}
    	\renewcommand{\arraystretch}{1.5}
    	\centering
    	\setlength{\tabcolsep}{0.8mm}
    	\begin{tabular}{|c|c|c|c|}
    		\hline
    		\multicolumn{4}{|c|}{$p = 0.02 ~,~ g_0 = 1.2$} \\
    		\hline
    		$q$ & 0.193 & 0.194 & 0.195 \\
    		\hline
    		$n_s$ & ~0.9566~ & ~0.9628~ & ~0.9689~ \\
    		\hline
    		$10^{62}r$ & 2.99 & 3.71 & 3.54 \\
    		\hline
    	\end{tabular}
    \end{minipage} 
    \hfill
    \begin{minipage}{0.33\linewidth}
    	\renewcommand{\arraystretch}{1.5}
    	\centering
    	\setlength{\tabcolsep}{0.8mm}
    	\begin{tabular}{|c|c|c|c|}
    		\hline
    		\multicolumn{4}{|c|}{$p = 0.02 ~,~ g_0 = 1.3$} \\
    		\hline
    		$q$ & 0.193 & 0.194 & 0.195 \\
    		\hline
    		$n_s$ & 0.9566 & 0.9628 & 0.9689 \\
    		\hline
    		$10^{62}r$ & 2.60 & 2.93 & 2.90 \\
    		\hline
    	\end{tabular}
    \end{minipage}

    \caption{ Predictions for the scalar spectral index $n_s$ and the 
tensor-to-scalar ratio $r$ for various combinations of $q$, $p$ and $g_0$.  The 
rest parameters are fixed according to \eqref{eq:parameter}, alongside 
the approximations $\rho_{r,B}T_B = 10^{-16}$(in Planck units) and $H_r \simeq 
H_{B+}$. }
    \label{tab:qpg}	
\end{table}

 Now we can proceed to the comparison of our results with observation data 
\cite{Aghanim:2018eyx}.  
In the thermal fluctuation case, the spectra index $n_s$  is determined solely 
by the parameter $q$, while the tensor-to-scalar ratio $r$ depends on various 
parameters. We start our investigation by examining the effect of the parameter 
$p$ and $g_0$, which determine the deformation of kinetic term from the 
canonical one, and hence the physics during the bounce phase. The other 
parameters are fixed according to \eqref{eq:parameter}, and the 
external parameters are taken to be $\rho_{r,B}T_B = 10^{-16}$ and $H_r \simeq 
H_{B+}$. In Table 
\ref{tab:qpg} we summarize the obtained observable predictions. In the above 
estimations we  have used $
z_{eq}=3400$ and $H_0=67.4~\rm{km~ s^{-1} Mpc^{-1}}$, i.e. $H_0=5.9\times 
10^{-61}$ in Planck units \cite{Aghanim:2018eyx}.

As we observe, the predicted tensor-to-scalar ratio $r$ is  extremely small 
and obviously well under the observational constraint $r < 0.007$ by orders of
magnitudes. As we discussed above, this result arises from the physics during 
the bounce phase: the conformal wavenumber at pivot scale $k_{\star} = 
0.05Mpc^{-1}$ is much smaller than the Hubble parameter $\mathcal{H}_{B-}$, 
which is of order of unity in inversely Planck length. Since, $r$ is  
suppressed by the combination $k/(-\mathcal{H}_{B-})$ in  
\eqref{eq:rthermalcase}, we can easily see why the obtained $r$-value is that 
small. This is the main result of the present work, and acts as a significant 
advantage of the proposed scenario, since it bypasses the main issue of usual 
matter bounce models, namely the  relatively high tensor-to-scalar ratio.

\begin{table}[h]

	\begin{minipage}{0.2\linewidth}
    	\renewcommand{\arraystretch}{1.5}
    	\centering
    	\setlength{\tabcolsep}{0.3mm}
    \begin{tabular}{|c|c|c|c|}
	\hline
	\multicolumn{4}{|c|}{$q = 0.193 ~,~ n_s = 0.9566$} \\
	\hline
	$H_r$ & $1.77 \times 10^{-4}$ & $1.77 \times 10^{-14}$ & $1.00 \times 10^{-42}$ \\
	\hline
	$\rho_{r,B}T_B$ & $7.95 \times 10^{-16}$ & $4.96 \times 10^{-28}$ & $2.12 \times 10^{-29}$ \\
	\hline
	$r$ & $4.40 \times 10^{-63}$ & $3.47 \times 10^{-64}$ & $1.25 \times 10^{-74}$ \\
	\hline
    \end{tabular}
   \end{minipage}
	\hfill
	\begin{minipage}{0.49\linewidth}
		\renewcommand{\arraystretch}{1.5}
		\centering
		\setlength{\tabcolsep}{0.3mm}
		\begin{tabular}{|c|c|c|c|}
	\hline
	\multicolumn{4}{|c|}{$q = 0.194 ~,~ n_s = 0.9628$} \\
	\hline
	$H_r$ & $1.77 \times 10^{-4}$ & $1.77 \times 10^{-14}$ & $1.00 \times 10^{-42}$ \\
	\hline
	$\rho_{r,B}T_B$ & $1.19 \times 10^{-16}$ & $7.60 \times 10^{-28}$ & $3.47 \times 10^{-29}$  \\
	\hline
	$r$ & $3.70 \times 10^{-62}$ & $2.90 \times 10^{-64}$ & $1.01 \times 10^{-74}$ \\
	\hline
    \end{tabular}
		\end{minipage}

	\begin{minipage}{1.0\linewidth}
		\renewcommand{\arraystretch}{1.5}
		\centering
		\setlength{\tabcolsep}{0.3mm}
		\begin{tabular}{|c|c|c|c|}
	\hline
	\multicolumn{4}{|c|}{$q = 0.195 ~,~ n_s = 0.9689$} \\
	\hline
	$H_r$ & $1.76 \times 10^{-4}$ & $1.76 \times 10^{-14}$ & $1.00 \times 10^{-42}$ \\
	\hline
	$\rho_{r,B}T_B$ & $1.83 \times 10^{-16}$ & $1.18 \times 10^{-27}$ & $5.70 \times 10^{-29}$  \\
	\hline
	$r$ & $3.12 \times 10^{-62}$ & $2.44 \times 10^{-64}$ & $8.34 \times 10^{-75}$ \\
	\hline
    \end{tabular}
	\end{minipage}
	
	\caption{ Predictions for the scalar spectral index $n_s$ and the 
tensor-to-scalar ratio $r$ for various combinations of $q$, $H_r$ and 
$\rho_{r,B}T_B$. The rest parameters are fixed according to equation 
\eqref{eq:parameter}. The parameters $H_r$ and $\rho_{r,B}T_B$ are 
suitably chosen    in order for the amplitude of the scalar spectrum to 
be equal  to the  values given by CMB observations, i.e. $A_s(K = 0.05Mpc^{-1}) 
= 2.1 \times 10^{-9}$.}
	\label{tab:qht}
	
\end{table}

Finally,   we investigate the effect of the ``external parameters'' 
$\mathcal{H}_r$ and $\rho_{r,B}T_B$ on the observables. $\mathcal{H}_r$ 
represents the energy scale of the radiation dominated phase, whereas 
$\rho_{r,B}T_B$ reflects the property of the thermal gas. Both     quantities 
are independent of the model parameters \eqref{eq:parameter}, and by 
studying their effect we can see that the CMB constraints can be satisfied 
within a vast range of parameters. Here, we summarize the results in Table 
\ref{tab:qht}.

\section{Conclusion and Discussions}
\label{sec:conclusion}
In this work, we investigated the bounce realization in the framework of DHOST 
cosmology. Although the basic scenario was developed in \cite{Ilyas:2020qja}, 
in the present investigation we focused on the detailed analysis of  
cosmological perturbations, and in particular on their relation with 
observables.

Firstly we chose  the involved functions and model parameters in order to 
acquire a desirable background, bouncing behavior followed by a fast-roll 
expansion, examining the involved matching conditions. Then, we proceeded to a 
detailed investigation of the scalar and tensor perturbations, during the 
Ekpyrotic contraction phase, the bounce phase, and finally during the fast-roll 
expansion phase, calculating the scalar and tensor power spectra and thus the 
spectral indices and the tensor-to-scalar ratio. Furthermore, we studied the  
proper initial conditions for these perturbations, incorporating three sets of 
such  conditions, namely perturbations generated by Ekpyrotic vacuum 
fluctuations, by matter vacuum fluctuations, and by thermal fluctuations.

The DHOST bounce scenario  with   cosmological perturbations  generated by 
thermal fluctuations proves to be the most efficient one. The corresponding 
predictions are in agreement with the observational bounds, and especially the 
tensor-to-scalar ratio $r$ is many orders of magnitude within the region 
allowed by Planck. The reason for this behavior is that the tensor spectrum generated in Ekpyrotic phase is blue, so the expression for $r$ contains the factor $\frac{k_{\star}}{\mathcal{H}_{B-}}^{n_t-1}$. Since $k_{\star}$ and $\mathcal{H}_{B-}$ are in astrophysical and Planck scale respectively, this factor suppress $r$ by magnitudes. Hence, the DHOST bounce scenario  bypasses the 
known disadvantage of most matter bounce realizations, namely the relatively 
high tensor-to-scalar ratio. 

Our analysis shows the DHOST coupling has negligible effect on the scalar and tensor power spectra, while the blue-tilted tensor spectrum in thermal initial condition case cam help to suppress the tensor-to-scalar ratio $r$, and these are the main conclusions of the present work. Hence, DHOST 
bounce cosmology could be further investigated. One such direction could be the 
examination of the two mechanisms that generate nearly scale-invariant scalar 
power spectra, namely the matter contraction and the thermal gas, in relation 
to the introduced extra parameters, i.e  the end of matter contraction period 
$\tau_m$ and the temperature of the thermal gas at the bounce point $T_B$, or 
  investigate whether the incorporation of other mechanisms, such as the 
curvaton one \cite{Enqvist:2001zp,Lyth:2001nq,Cai:2011zx}, could lead to scale invariance. Additionally, since our  perturbation  analysis  in the 
Ekpyrotic and fast-roll expansion phases will remain valid for bounce 
cosmologies with multiple bounce phases, as long as the total bouncing duration 
is negligible compared to the whole cosmological process (since the bounce 
phase, if lasting for a short time,   contributes only an amplification factor 
to the scalar power spectrum), we could study the extension of the scenario at 
hand  to the multiple bounce cosmology. Finally, phenomenology of DHOST bounce cosmology is worth investigating. One possibility is to study the reheating and preheating process, like that in \cite{Cai:2011ci}. Another possibility is that, since a phase of NEC violation can produce observable signals on gravitational waves in inflationary cosmology \cite{Tahara:2020fmn,Cai:2020qpu}, it would be interesting to study the imprint on primordial gravitational waves from our model(especially from the NEC violated bounce phase). Such studies lie beyond the scope of 
the present work and are left for future projects.

		\acknowledgments
	We are grateful to Robert Brandenberger,  Qianhang Ding, Damien Easson, Xian 
Gao, Misao Sasaki, Dong-Gang Wang, Yi Wang, Zhi-bang Yao, Yong Cai and Siyi Zhou for 
stimulating discussions. 
This work is supported in part by the NSFC (Nos. 11722327, 11961131007, 11653002, 11847239, 11421303), by the CAST Young Elite Scientists Sponsorship (2016QNRC001), by the National Youth Talents Program of China, by the Fundamental Research Funds for Central Universities, by CAS project for young scientists in basic research YSBR-006, by the China Scholarship Council (CSC No.202006345019),by the USTC Fellowship for international students under the ANSO/CAS-TWAS scholarship, and by GRF Grant 16304418 from the Research Grants Council of Hong Kong. YZ would like to thank the ICRANet for their hospitality during his visit. 
All numerical calculations are operated on the computer clusters {\it LINDA \& 
JUDY} in the particle cosmology group at USTC.

	\appendix
	\section{A Brief Introduction to DHOST theory}
	\label{app:DHOST}

		In this Appendix, we present a brief introduction to the DHOST theory. 
DHOST theories are defined to be the maximal set of scalar-tensor theories in 
four dimensional space-time that contain at most three powers of second 
derivatives of the scalar field $\pi$, while propagating at most three degrees 
of freedom. Note that the Galileon theory is the specific sub-class of DHOST 
theories where the equation of motion for the scalar field remains second 
order. 
	
	The most general DHOST action involving up to cubic powers of second 
derivatives of the scalar field $\pi$ can be written as 
	\begin{equation}
		\label{dhost32}
		S[g,\pi]  = \int d^4x \sqrt{-g} \Big[ h_2(\pi, X) R +  C^{\lambda \nu 
\rho \delta}_{(2)} \pi_{\lambda \nu} \pi_{\rho \delta}   
		  + h_3(\pi, X) G_{\lambda \nu}  \pi^{\lambda \nu} +C^{\lambda \nu \rho 
\delta \alpha \beta}_{(3)} \pi_{\lambda \nu} \pi_{\rho \delta} \pi_{\alpha 
\beta} \Big] ~.
	\end{equation} 
	The tensors $C_{(2)}$ and $C_{(3)}$ represent the most general tensors 
constructed with the metric $g_{\lambda \nu}$ as well as the first derivative of 
the scalar field,  which is denoted as $\pi_{\lambda} \equiv \nabla_{\lambda} 
\pi$. The symbol $\pi_{\lambda \nu}$ denotes the second derivative $\pi_{\lambda 
\nu} \equiv \nabla_{\lambda} \nabla_{\nu} \pi$, and the canonical kinetic  term 
$X$ is defined as $X \equiv \frac{1}{2} \nabla^{\nu} \pi \nabla_{\nu} \pi$. 
Exploiting the symmetries of  $C_{(2)}$ and  $C_{(3)}$, one can reformulate 
 \eqref{dhost32} through
	\begin{equation}\nonumber
		C^{\lambda \nu \rho \delta}_{(2)} \pi_{\lambda \nu} \pi_{\rho \delta} +  
 C^{\lambda \nu \rho \delta \alpha \beta}_{(3)} \pi_{\lambda \nu} \pi_{\rho 
\delta} \pi_{\alpha \beta} \equiv \sum_{i=1}^5 a_i L_i^{(2)} + \sum_{j=1}^{10} 
b_j L_j^{(3)} ~,
	\end{equation}
	where $a_i$ and $b_i$ depend on $\pi$ and $X$. Since we only need the 
quadratic DHOST terms, in our model we consider all $b_j$ terms to vanish.
	
	The involved Lagrangians      are defined as
	\begin{align*} 
		\label{eq:L_i^2}
		& L_1^{(2)} = \pi_{\mu \nu} \pi^{\mu \nu} ~,~ L_2^{(2)} = (\Box \pi)^2 ~,~ L_3^{(2)} =  (\Box \pi)\pi^{\mu}\pi_{\mu \nu} \pi^{\nu} ~, \nonumber\\ 
		& L_4^{(2)} = \pi_{\mu}\pi^{\mu \rho}\pi_{\rho \nu}\pi^{\nu} ~,~ L_5^{(2)} = (\pi^{\mu}\pi_{\mu \nu}\pi^{\nu})^2 ~,
	\end{align*}
	and to make the theory not propagating the ghost degree of freedom, the kinetic matrix of the action \eqref{dhost32} should be degenerate, hence the 
form of $a_i$'s are severely constrained. There are six possible combinations of 
$a_i$'s which can give a healthy action without ghost, and for our purposes we 
first concentrate on the type $^{(2)} N$\textendash$II$ DHOST theory, where 
there are three free functions $h_2$, $a_4$ and $a_5$, and the others are 
constrained by
	\begin{equation}
		\label{eq:dhostconstraint}
		a_2 = -a_1 = \frac{h_2}{2X} ~,~ a_3 = \frac{h_2 - 2 X h_{2X}}{2 X^2} ~.
	\end{equation}
As we can see from  \eqref{eq:dhostconstraint}, the action in 
\eqref{eq:Action} can be seen as the merge of the type $^{(2)} N$\textendash$II$ DHOST action \eqref{eq:LD} with Horndeski action $R/2 + K(\phi,X) + Q(\phi,X)\Box \phi$
\cite{Ilyas:2020qja}.

Finally we mention that, our model \eqref{eq:Action} can also be viewed as the combination of Horndeski action $K + Q\Box \phi$ and a type $^{(2)}N-I$ DHOST action $\mathcal{L}_D + R/2$. The type $^{(2)}N-I$ DHOST theory is defined through the three free functions $h_2$, $a_1$ and $a_3$, with the constraints 
\begin{equation}
	a_2 = -a_1 \neq \frac{h_2}{2X} ~,
\end{equation}
\begin{align}
	a_4  \nonumber = \frac{1}{8(h_2 + 2a_1X)^2} & \{ 4h_2 [ 3(a_1 + h_{2X})^2 - 2a_3h_2 ] - 4a_3X^2 (a_3h_2 - 8a_1h_{2X}) \\
	& -8X (3a_1a_3h_2 -8a_1^2h_{2X} - 4a_1h_{2X}^2 -4a_1^3 - a_3h_2h_{2X}) \} ~,
\end{align}
\begin{equation}
	a_5 = \frac{1}{2(h_2 + 2a_1X)^2}(a_1 + a_3X + h_{2X}) [a_1(a_1 - 3a_3X + h_{2X}) - 2a_3h_2] ~.
\end{equation}
One can verify that the action $\mathcal{L}_D + R/2$ belongs to the type $^{(2)}N-I$ DHOST theory by taking
\begin{equation}
	h_2 = \frac{1+f}{2} ~,~ a_1 = -\frac{f}{2X} ~,~ a_3 = \frac{f - 2Xf_{X}}{2X^2} ~. 
\end{equation}

As the action $K + Q\Box \phi$ won't change the degenerate condition and thus the classification, our model can be classified as the type $^{(2)}N-I$ DHOST theory. However, since the action $\mathcal{L}_D$ doesn't contribute to the background dynamics as well as observable quantities like $n_s$ and $r$, we prefer to classify $\mathcal{L}_D$ as the DHOST coupling.

	\section{Dynamics of Scalar Perturbations during Bounce Phase}
	\label{app:scalarbounce}
	
	In this Appendix we briefly discuss the dynamics of scalar perturbations 
during the bounce phase. The corresponding dynamical equation is 
\eqref{eq:MSeqSb}, whose general solution is the parabolic cylinder function. 
However,    the solution can be approximated 
by exponential functions as:
	\begin{equation}
		\label{eq:vkbexp}
		v_k^b(\tau) \simeq b_{s,3}(k)e^{\int \omega_k(\tau) d\tau}  + 
b_{s,4}(k)e^{-\int \omega_k(\tau) d\tau} ~,~ \omega_k^2(\tau) = 
\frac{4\tau^2}{T^4} + \frac{2}{T^2} - c_s^2 k^2 ~,
	\end{equation}
	where we have omitted the $\Gamma$ dependent term in the definition of 
$\omega_k$.  Furthermore, the $c_s^2k^2$ term is subdominant in the expression 
of $\omega_k$, therefore the dynamics of scalar perturbations is almost 
scale-independent. The overall effect on $v_k$ during the bounce phase is an 
amplification independent of $k$, with the amplification factor $\mathcal{F}_s$ 
being 
	\begin{equation}
		\mathcal{F}_s \simeq \exp{ \left( \int_{\tau_{B-}}^{\tau_{B+}} \omega_k d\tau \right)} \simeq e^{11} = 6 \times 10^4 ~.
	\end{equation}
	
	Additionally, we can   solve the differential equation 
\eqref{eq:MSeqSb} numerically, too. We set $T = 0.5$, $\tau_{B-} = \tau_{B+} = 
1$,   we neglect $c_s^2k^2$, and we choose two sets of ``initial conditions'', 
namely  $v_k(0) = 1 ~,~ v_k^{\prime}(0) = 0$ and $v_k(0) = 0 ~,~ 
v_k^{\prime}(0) = 1$. We solve \eqref{eq:MSeqSb} numerically and we compare the 
results with the semi-analytic solution \eqref{eq:vkbexp}. 
	
	\begin{figure}[htp]
		\centering
		\includegraphics[width=.4\textwidth]{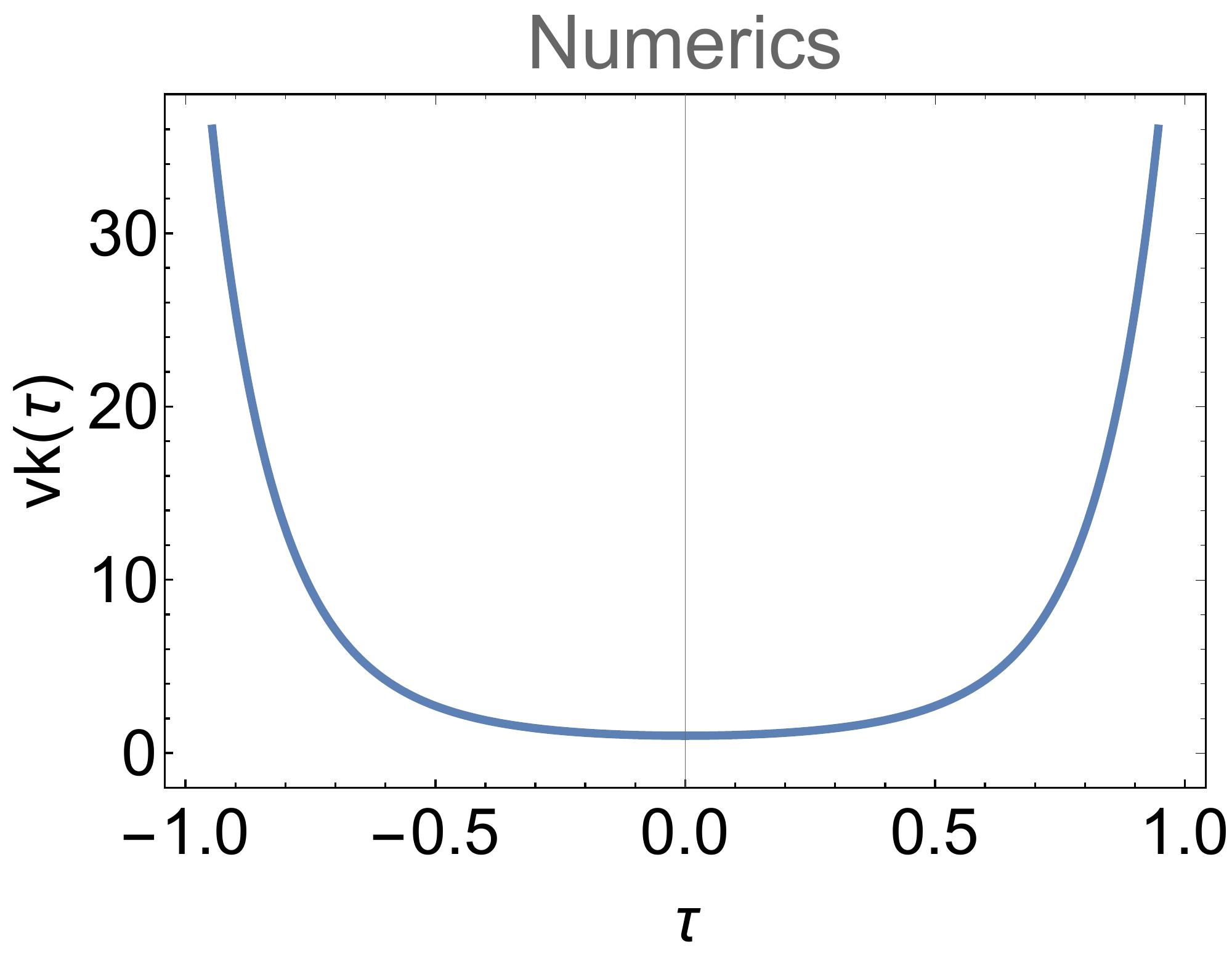} \quad 
		\includegraphics[width=.42\textwidth]{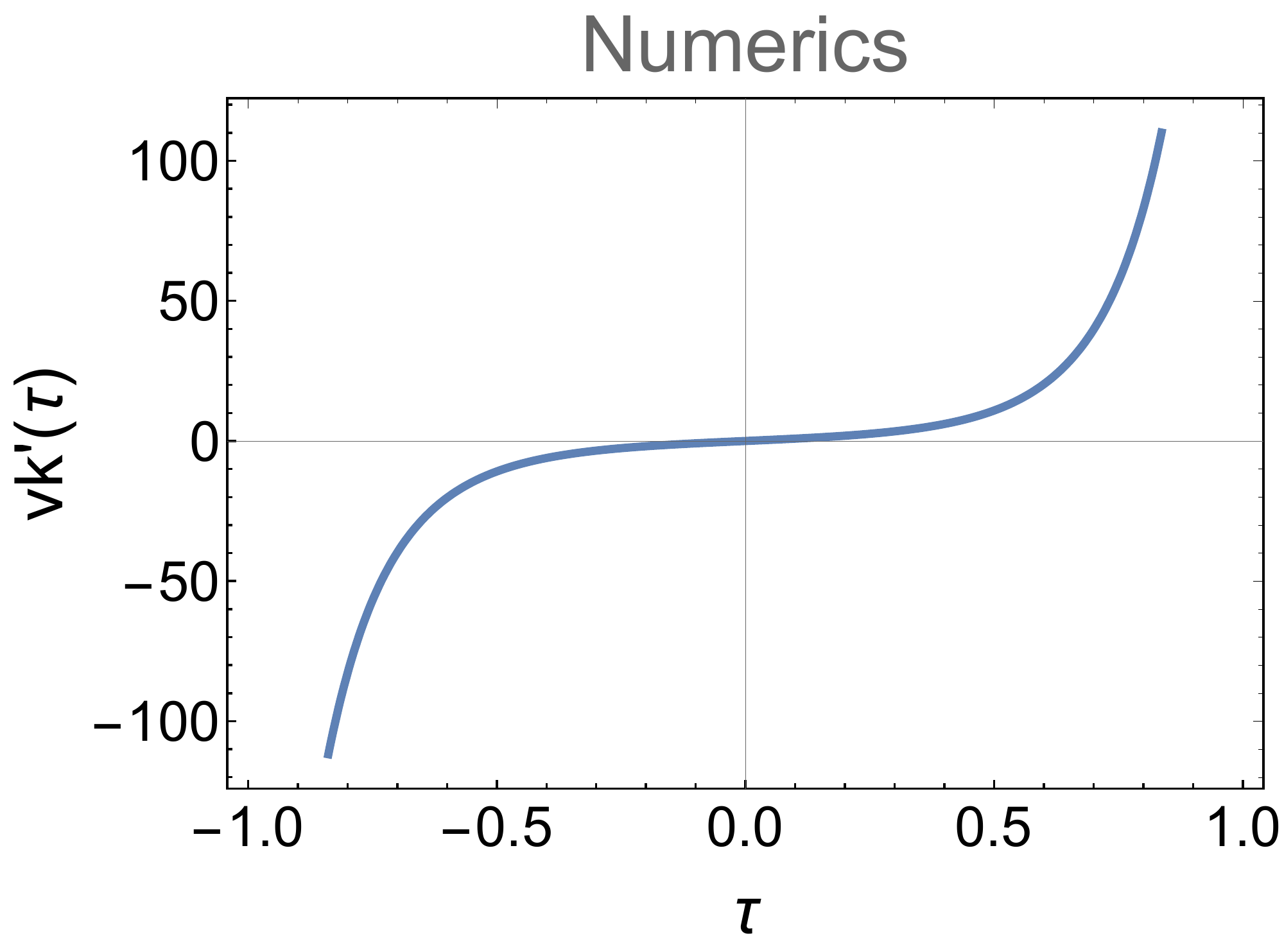} \\
		\includegraphics[width=.4\textwidth]{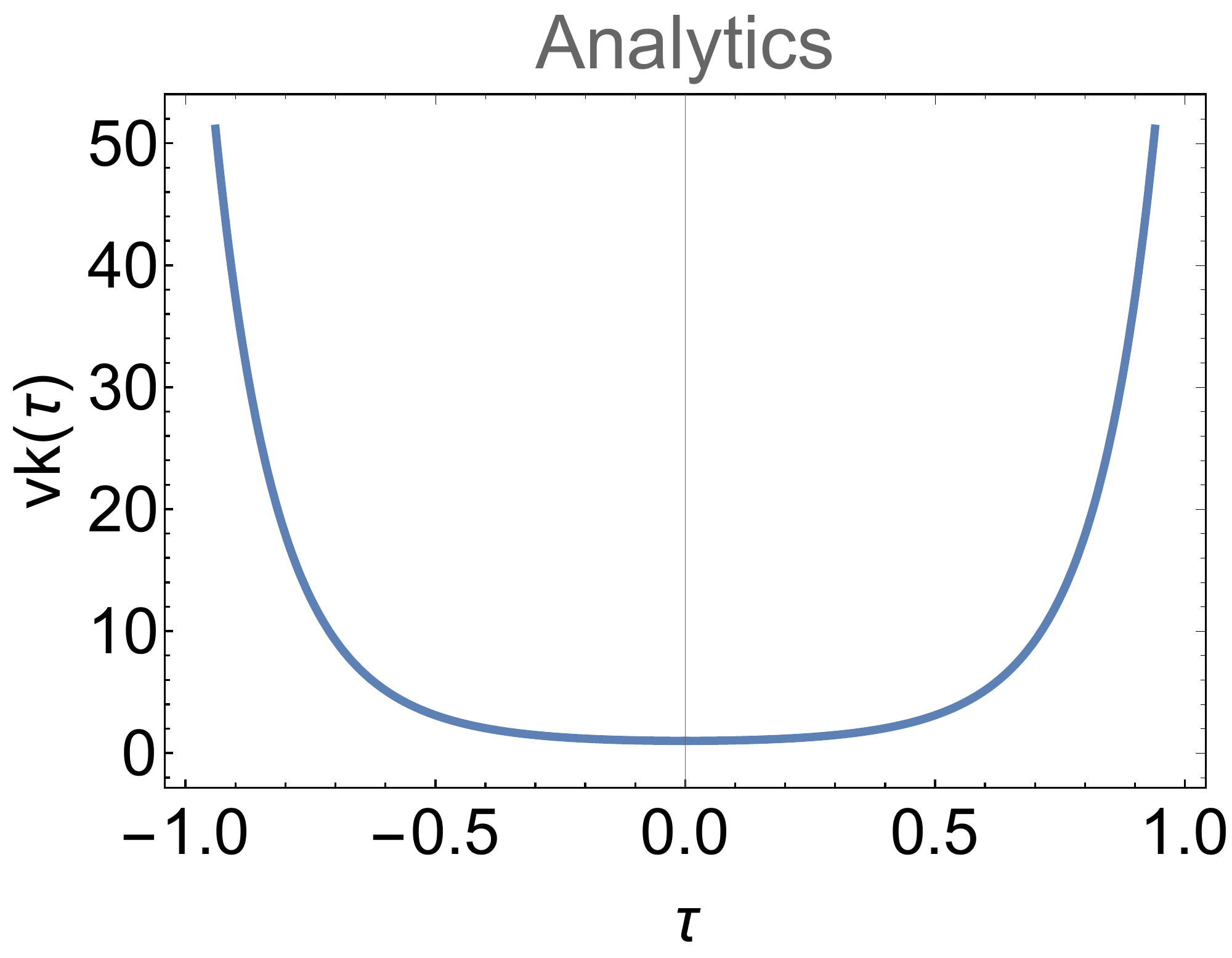} \quad
		\includegraphics[width=.42\textwidth]{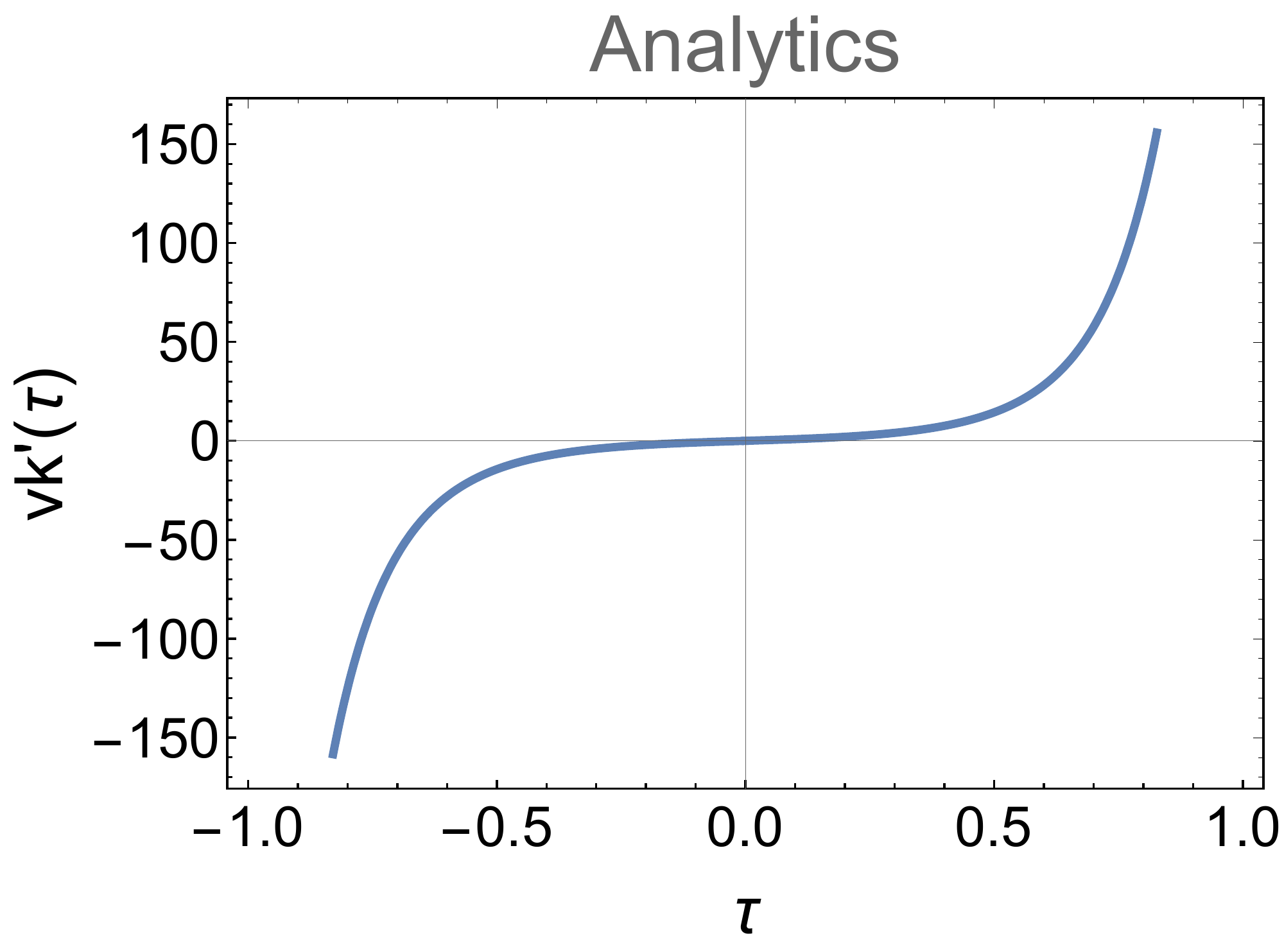}
		\caption{
		Comparison between the numerical solution to equation \eqref{eq:MSeqSb} 
(upper graphs) and its semi-analytical approximation 
\eqref{eq:vkbexp} (lower graphs) during the bounce phase. The left and right 
graphs   show $v_k$ and $v_k^{\prime}$, respectively. We have imposed 
initial conditions $v_k(0) = 1$ and $v_k^{\prime}(0) = 0$.  }
		\label{fig:vkbcompare1}
	\end{figure}
	
	\begin{figure}[htbp]
	\centering
		\includegraphics[width=.4\textwidth]{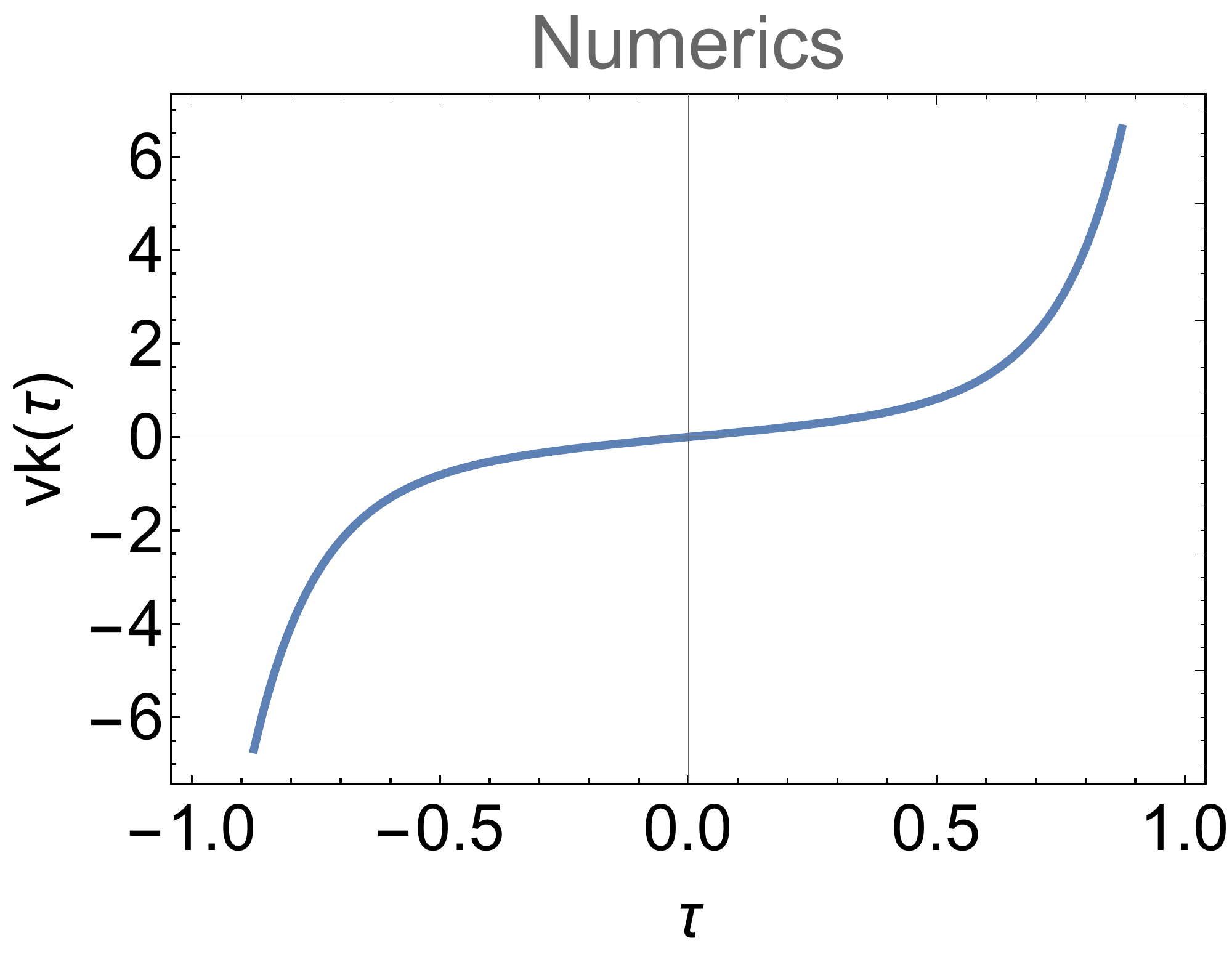} \quad 
		\includegraphics[width=.4\textwidth]{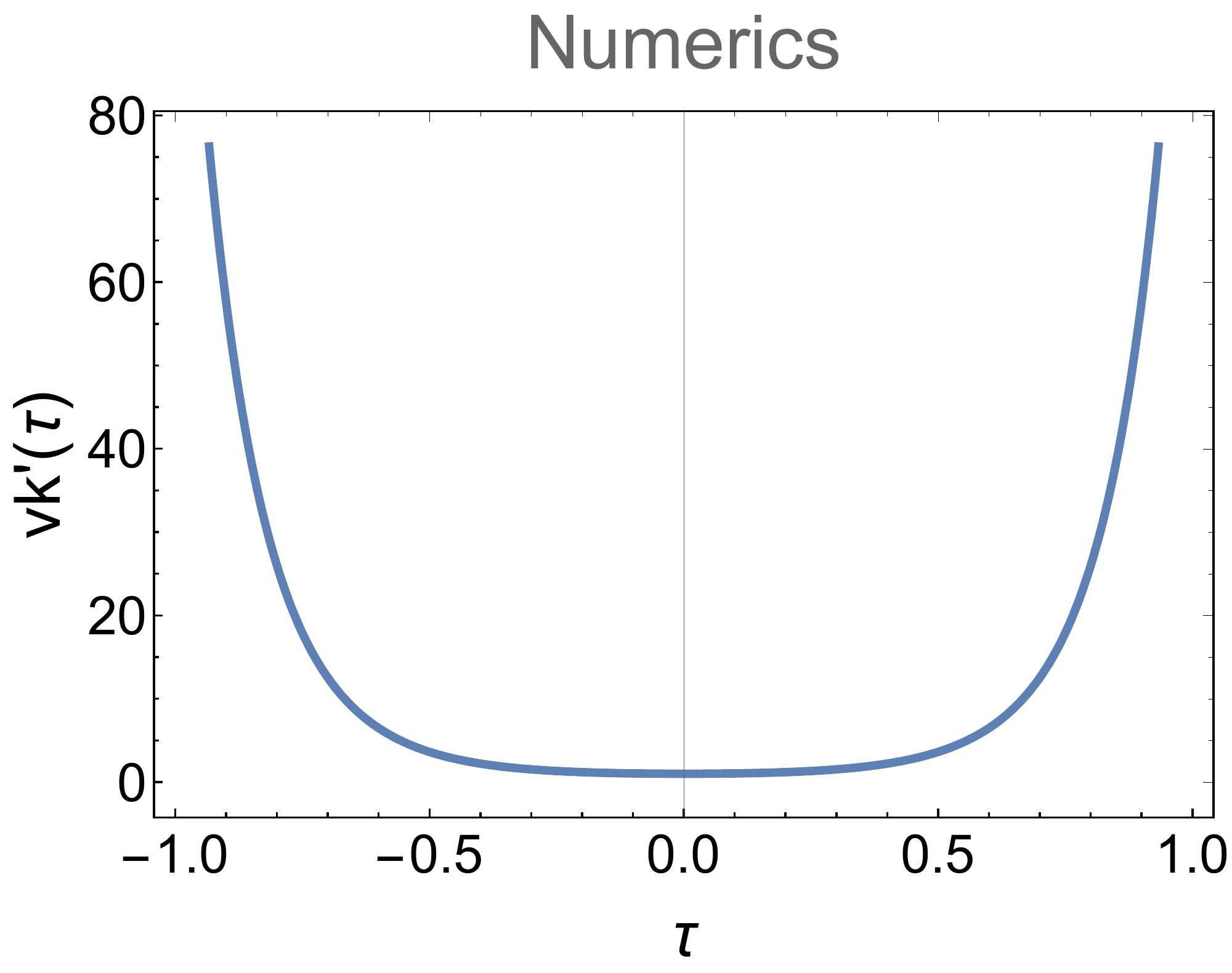} \\
		\includegraphics[width=.4\textwidth]{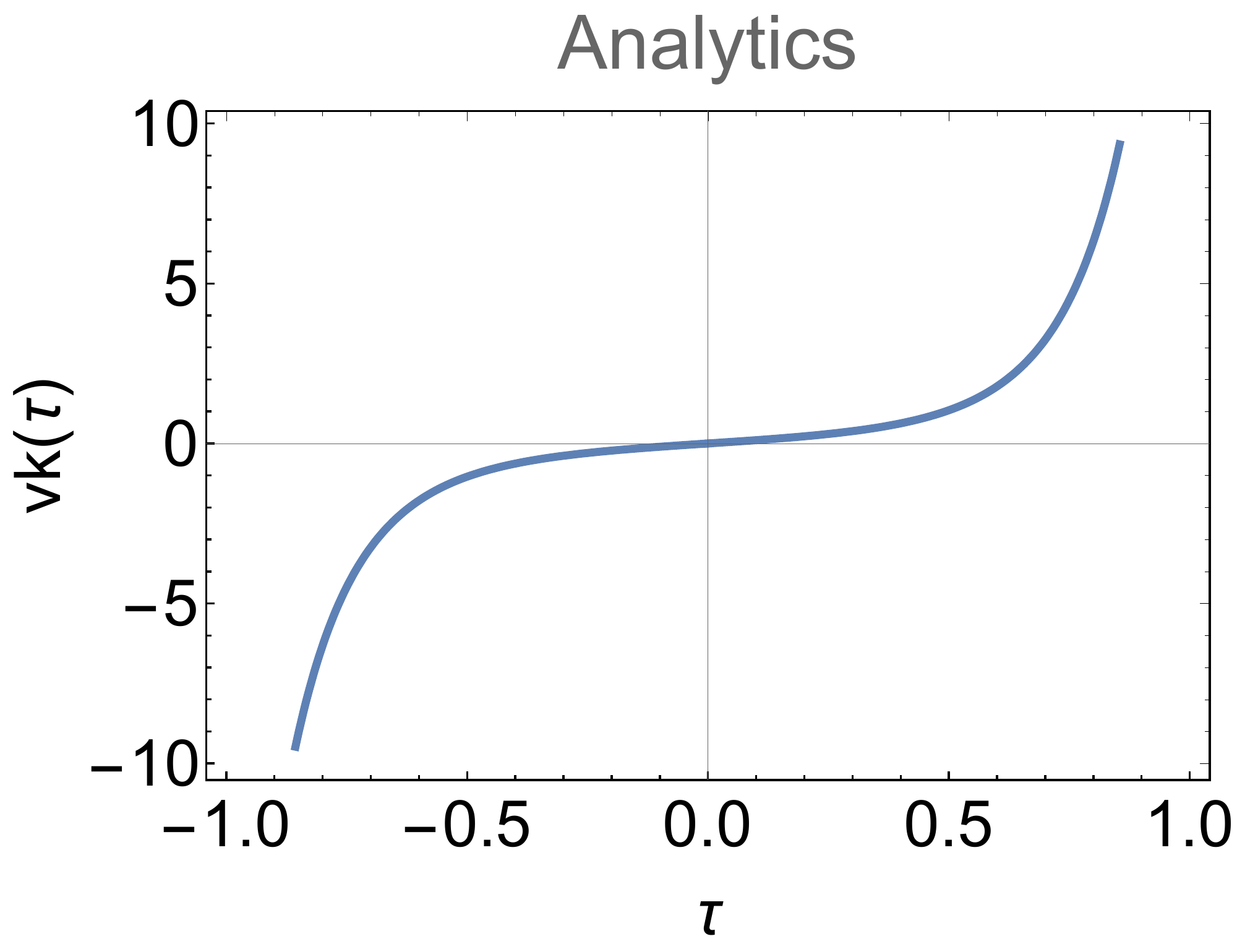} \quad
		\includegraphics[width=.4\textwidth]{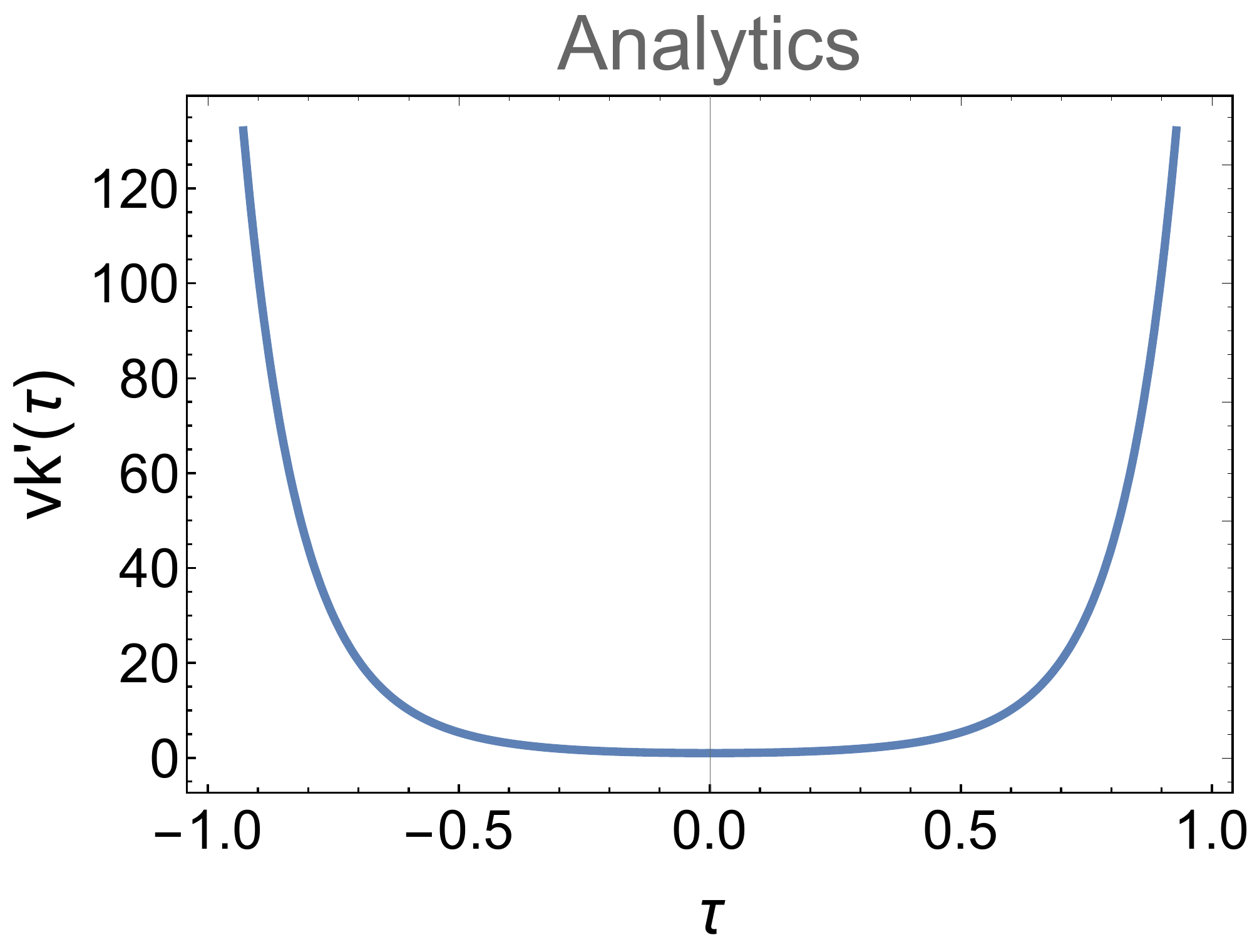}
		\caption{
		Comparison between the numerical solution to equation \eqref{eq:MSeqSb} 
(upper graphs) and its semi-analytical approximation 
\eqref{eq:vkbexp} (lower graphs) during the bounce phase. The left and right 
graphs   show $v_k$ and $v_k^{\prime}$, respectively. We have imposed 
initial conditions  $v_k(0) = 0$, 
$v_k^{\prime}(0) = 1$. }
		\label{fig:vkbcompare2}
	\end{figure}
	
	For $v_k(0) = 1 ~,~ v_k^{\prime}(0) = 0$, the corresponding coefficients in 
\eqref{eq:vkbexp} are $b_{s,3} = b_{s,4} = \frac{1}{2}$.  In Fig. 
\ref{fig:vkbcompare1}   we present the results. Similarly, for $v_k(0) = 0 ~,~ 
v_k^{\prime}(0) = 1$, the corresponding coefficients in \eqref{eq:vkbexp} are 
$b_{s,3} = -b_{s,4} = \frac{1}{2\omega_k(0)}$, and in Fig.  
\ref{fig:vkbcompare2}    we depict the resulting behavior. In both cases, we 
deduce that the numerical and analytical results are in perfect agreement, 
apart from an insignificant multiplication factor     of the order 
$\mathcal{O}(1)$. Hence, equation \eqref{eq:vkbexp} indeed provides a good 
approximation to the solution of \eqref{eq:MSeqSb}, and thus we shall use 
equation \eqref{eq:vkbexp} for the following calculations.

	\section{Matching Conditions for Scalar Perturbations}
	\label{app:scalarmatch}

	In this Appendix  we derive the matching conditions for scalar 
perturbations across the transition surface. The first matching surface is $\tau 
= \tau_{B-}$, and we shall match equation \eqref{eq:vkcontraction} with 
\eqref{eq:vkb}. At the time $\tau = \tau_{B-}$, the dominant part of 
\eqref{eq:vkcontraction} is the Bessel $Y$ function, therefore we may take
	\begin{equation}
		v_k^c(\tau) \simeq b_{s,2} \sqrt{\tau_- - \tau} Y_{\nu_c}[k(\tau_- - \tau)] \simeq - b_{s,2} \frac{\Gamma(\nu_c)}{\pi} \left( \frac{2}{k} \right)^{\nu_c} (\tau_- - \tau)^{\frac{1}{2} - \nu_c} ~,
	\end{equation}
	where $\Gamma(x)$ is the gamma function. Hence we obtain
	\begin{equation}
		v_k^c(\tau_{B-}) \simeq - b_{s,2} \frac{\Gamma(\nu_c)}{\pi} \left( \frac{2}{k} \right)^{\nu_c} (\tau_- - \tau_{B-})^{\frac{1}{2} - \nu_c} ~, 
	\end{equation}
	\begin{equation}
		\frac{dv_k^c}{d\tau}(\tau_{B-}) \simeq b_{s,2} \frac{\Gamma(\nu_c)}{\pi} \left(\frac{1}{2} - \nu_c \right) \left( \frac{2}{k} \right)^{\nu_c} (\tau_- - \tau_{B-})^{-\frac{1}{2} - \nu_c} ~.
	\end{equation}
	
	As we analyzed in Section \ref{sec:ptbounce}, the dynamics of scalar 
perturbations during the bounce phase can be simplified as
	\begin{equation}
		\label{eq:vkbapp3}
		v_k^b(\tau) \simeq b_{s,3}e^{W_k(\tau)} + b_{s,4}e^{-W_k(\tau)} ~,~ \frac{dv_k^b}{d\tau}(\tau) = \omega_k \left( b_{s,3}e^{W_k(\tau)} - b_{s,4}e^{W_k(\tau)} \right) ~,
	\end{equation}
	with the help of the definition
	\begin{equation}
		W_k(\tau) \equiv \int_{\tau_{B-}}^{\tau} \omega_k(x)dx ~ \to ~ W_k^{\prime}(\tau) = \omega_k(\tau) ~. 
	\end{equation}
	Note that here the lower limit of integration in $W_k(\tau)$ is taken as 
$\tau_{B-}$, different from the case of Appendix \ref{app:scalarbounce} above 
where the lower limit was 0. Since the change of the integration range can
correspond to a redefinition of $b_{s,3}$ and $b_{s,4}$,  
equation \eqref{eq:vkbapp3} remains valid. Finally, for convenience  we 
introduce the abbreviation
	\begin{equation}
		\omega_{k-} \equiv \omega_k(\tau_{B-}) ~,~ \omega_{k+} \equiv \omega_k(\tau_{B+}) ~.
	\end{equation}
	
	The perturbation $v_k$ at the transition surface $\tau = \tau_{B-}$ can be 
found by  directly substituting the time into equation \eqref{eq:vkbapp3}. 
However, at $\tau = \tau_{B+}$ we observe that the $b_{s,4}$ term in 
transition surface $\tau = \tau_{B+}$ can be omitted: during the bounce phase, 
the $b_{s,4}$ term will be suppressed by a factor $\mathcal{F}_s$, while the 
$b_{s,3}$ term acquires an amplification by the same factor. 
The factor 
$\mathcal{F}_s$ is generally large\footnote{The fact that 
$\mathcal{F}_s$ 
is large arises from the tachyonic instability during the bounce phase, 
which can be applied to the bounce scenario to suppress the tensor-to-scalar 
ratio \cite{Cai:2011zx} and improve the efficiency of preheating process 
\cite{Cai:2011ci}.} (for example in the numerical case we have $\mathcal{F}_s 
\sim \mathcal{O}(10^4)$) and depends on $k$. Hence, the $b_{s,4}$ term will be 
negligible at $\tau = \tau_{B+}$, and we have
	\begin{equation}
		v_k^b(\tau_{B+}) \simeq b_{s,3} \mathcal{F}_s ~,~ \frac{dv_k^b}{d\tau} (\tau_{B+}) \simeq b_{s,3}\omega_{k+} \mathcal{F}_s ~,
	\end{equation}
	where we have used the fact $\mathcal{F}_s \simeq W_k(\tau_{B+})$.
	
	Finally, we use the small-argument expansion of the Bessel function to 
rewrite equation \eqref{eq:vke} at $\tau \to \tau_{B+}$ as
	\begin{equation}
		v_k^e(\tau) \simeq b_{s,5} \sqrt{\tau - \tau_+} +  \frac{2}{\pi} \left[ \ln \frac{k(\tau - \tau_+)}{2} + \gamma_E \right] b_{s,6} \sqrt{\tau - \tau_+} ~,
	\end{equation}
	where $\gamma_E \simeq 0.58$ is the Euler-Masheroni constant. At the transition surface $\tau = \tau_{B+}$, we have
	\begin{equation}
		v_k^e(\tau_{B+}) \simeq b_{s,5}\sqrt{\tau_{B+} - \tau_+} + \frac{2}{\pi} \left[ \ln \frac{k(\tau_{B+} - \tau_+)}{2} + \gamma_E \right] b_{s,6} \sqrt{\tau_{B+} - \tau_+} ~,
	\end{equation}
	\begin{equation}
		\frac{dv_k^e}{d\tau}(\tau_{B+}) = \frac{1}{2\sqrt{\tau_{B+} - \tau_+}} \left[ b_{s,5} + \frac{2b_{s,6}}{\pi} \left( 2 + \gamma_E + \ln \frac{k(\tau_{B+} - \tau_+)}{2} \right) \right] ~.
	\end{equation}	
	Equating $v_k$ and $v_k^{\prime}$ at $\tau = \tau_{B-}$ and eliminating 
$b_{s,4}$, we have
	\begin{equation}
		\label{eq:bs3}
		b_{s,3} = \frac{b_{s,2}\Gamma(\nu_c) \left[ 1 - 2\nu_c - 2(\tau_- - ,
\tau_{B-})\omega_{k-} \right]}{2^{2-\nu_c} \pi  k^{\nu_c} (\tau_- - 
\tau_{B-})^{\frac{1}{2} + \nu_c} \omega_{k-}} ~,
	\end{equation}
	where we have  used that $W_k(\tau_{B-}) = 0$ by definition. 
Determining $b_{s,3}$ is sufficient to extract all information about $v_k$ at 
$\tau = \tau_{B+}$. Moreover,    we should acquire $b_{s,5}$ and $b_{s,6}$ in 
equation \eqref{eq:vke}. Hence, equating $v_k$ and $v_k^{\prime}$ at $\tau = 
\tau_{B+}$  we obtain
	\begin{eqnarray}
		\label{eq:bs5}
&&		b_{s,5} = \frac{b_{s,3} \mathcal{F}_s}{\sqrt{\tau_{B+} - \tau_+}} 
\left\{ 1 -  \left[ (\tau_{B+} - \tau_+)\omega_{k+} - \frac{1}{2} \right] \left[ 
\ln \frac{k(\tau_{B+} - \tau_+)}{2} + \gamma_E \right] \right\},
\\
		\label{eq:bs6}
&&		b_{s,6} = \frac{\pi}{4}\mathcal{F}_s b_{s,3} \frac{2(\tau_{B+} - 
\tau_+) \omega_{k+} - 1}{\sqrt{\tau_{B+} - \tau_+}} ~.
	\end{eqnarray}
	
	Before proceeding, we replace the combination  $(\tau_{B+} - \tau_+)$ 
through the Hubble parameter $\mathcal{H}_{B+}$, which represents the energy 
scale at the end of the bounce phase, since from   \eqref{eq:tau+}  we have 
$\tau_{B+} - \tau_+ = \frac{1}{2\mathcal{H}_{B+}}$. Combining 
\eqref{eq:bs3}, \eqref{eq:bs5} and \eqref{eq:bs6}, we can finally express the 
coefficients $b_{s,5}$, $b_{s,6}$ in terms of the initial condition $b_{s,2}$ as
	\begin{eqnarray}
		\label{eq:bs2tobs5}    
	&&	b_{s,5} = \frac{\mathcal{F}_s\Gamma(\nu_c)  \left[ 1 - 2\nu_c - 
2(\tau_- - \tau_{B-})\omega_{k-} \right]}{2^{\frac{5}{2}-\nu_c}\pi  k^{\nu_c} 
(\tau_- - \tau_{B-})^{\frac{1}{2} + \nu_c} \omega_{k-} \mathcal{H}_{B+}^{-1/2}} 
\left[ 2 - \left(\frac{\omega_{k+}}{\mathcal{H}_{B+}} - 1 \right) \ln \frac{k 
e^{\gamma_E}}{4\mathcal{H}_{B+}} \right] b_{s,2} ~,
\\
		\label{eq:bs2tobs6}
	&&	b_{s,6} = \frac{\mathcal{F}_s\Gamma(\nu_c)}{2^{\frac{7}{2}-\nu_c} 
k^{\nu_c}} \frac{ 1 - 2\nu_c - 2(\tau_- - \tau_{B-})\omega_{k-}}{(\tau_- - 
\tau_{B-})^{\frac{1}{2} + \nu_c} \omega_{k-}} \left[ 
\frac{\omega_{k+}}{\sqrt{\mathcal{H}_{B+}}} - \sqrt{\mathcal{H}_{B+}} \right] 
b_{s,2} ~.
	\end{eqnarray}
	
Expressions  \eqref{eq:bs2tobs5} and \eqref{eq:bs2tobs6} can be further 
simplified. Since we have a short bounce, $\mathcal{H}_{B+}$ and 
$\mathcal{H}_{B-}$ will have a relatively small value; in the specific case used 
in the main text of this manuscript, they are of the order of $10^{-4}$, 
thus $1/\mathcal{H}$ will be relatively large. On the other hand, for Ekpyrotic 
matter field we have $0<\nu_c<1/2$ since $0<q<1/3$, and $\omega_{k+}$ is of the
order of $\mathcal{O}(10)$ (in our case $\omega_{k+} \simeq 8.5$). Hence, we 
can neglect the term $1-2\nu_c$ and $\mathcal{H}_{B+}/\omega_{k+}$, and the 
final   expressions for $b_{s,5}$ and $b_{s,6}$ are
	\begin{equation}
		\label{eq:bs5final}    
		\frac{b_{s,5}}{b_{s, 2}} = \frac{\mathcal{F}_{s} \Gamma\left(\nu_{c}\right)\left(1-2\nu_c\right)^{\frac{1}{2}-\nu_{c}}}{4^{1-\nu_{c}}\pi}\left(\frac{-\mathcal{H}_{B-}}{k}\right)^{\nu_{c}} \left(  \frac{w_{k+}}{\sqrt{-\mathcal{H}_{B-} \mathcal{H}_{B+}}}\ln \frac{ke^{\gamma_E}}{4 \mathcal{H}_{B+}} - \sqrt{\frac{4\mathcal{H_{B+}}}{-\mathcal{H_{B-}}}} \right) ~,
	\end{equation}
	and
	\begin{equation}
		\label{eq:bs6final}    
		\frac{b_{s,6}}{b_{s, 2}} = -\frac{\mathcal{F}_s \Gamma(\nu_c) (1-2\nu_c)^{\frac{1}{2}-\nu_c}}{2^{3-2\nu_c}} \left( \frac{-\mathcal{H_{B-}}}{k} \right)^{\nu_c} \frac{\omega_{k+}}{\sqrt{-\mathcal{H_{B-}} \mathcal{H_{B+}} }} ~,
	\end{equation}
	where we have used that
	\begin{equation}
		\tau_- - \tau_{B-} = -\frac{q}{(1-q)\mathcal{H_{B-}}} ~.
	\end{equation}
		Equations \eqref{eq:bs5final} and \eqref{eq:bs6final} are the final 
results of this Appendix.

	\section{Matching Conditions for Tensor Perturbations}
	\label{app:tensormatch}
	
	In this Appendix we first deduce the matching conditions for tensor 
perturbations across the transition surface, and then we apply it to our model. 
The starting point is the MS equation \eqref{eq:MSeqtensor}, which can be 
rewritten as
	\begin{equation}
		\label{eq:MSeqtensor1st}
		\frac{d}{d\tau} \left( \mu_k^{\prime} a - \mu_k a^{\prime} \right) = - c_T^2 k^2 a \mu_k ~.
	\end{equation}
	Only for this Appendix   we define the auxiliary functions 
	\begin{equation}
		g(\tau) = \mu_k^{\prime}(\tau) a(\tau) - \mu_k(\tau) a^{\prime}(\tau) ~,~ M(\tau,\epsilon_1,\epsilon_2) = \int_{\tau - \epsilon_2}^{\tau + \epsilon_1} [- c_T^2 k^2 a(\tau) \mu_k(\tau)] d\tau ~,
	\end{equation}
	and thus when we integrate equation \eqref{eq:MSeqtensor1st} across one 
transition surface, e.g. $\tau = \tau_{B-}$, we obtain 
	\begin{equation}
		\label{eq:gM}
		g(\tau_{B-} + \epsilon_1) - g(\tau_{B-} - \epsilon_2) = M(\tau_{B-},\epsilon_1,\epsilon_2) ~.
	\end{equation}
	
	From the finiteness of $a(\tau)$ and $\mu_k(\tau)$, it is easy to see that
	\begin{equation}
		\lim_{\epsilon_1 \to 0 , \epsilon_2 \to 0} M(\tau_{B-},\epsilon_1,\epsilon_2) = 0 ~,
	\end{equation}
	and therefore
	 $g(\tau)$ should be finite near the transition surface, 
otherwise the limiting value of $M$ would be a finite number.
Then integrating \eqref{eq:gM} we have
	\begin{equation}
		\label{eq:muka}
		\left( \frac{\mu_k}{a} \right)(\tau_{B-} + \lambda) -  \left( 
\frac{\mu_k}{a} \right)(\tau_{B-} - \lambda) = \int_{\tau_{B-} - 
\lambda}^{\tau_{B-} + \lambda} [a(\tau)^2g(\tau)] d\tau ~.
	\end{equation}
	As we can see, in the limit $\lambda \to 0$   the 
right-hand-side of \eqref{eq:muka} will vanish due to the finiteness of 
$a(\tau)$ and $g(\tau)$. Hence, $\mu_k/a$ is continuous at $\tau = \tau_{B-}$, 
and alongside  the continuity of $a(\tau)$  we deduce that $\mu_k(\tau)$ 
is continuous at $\tau = \tau_{B-}$.
	Finally,  looking at \eqref{eq:gM}, and following the same argument, we 
find that  $g(\tau)$ is continuous at $\tau = \tau_{B-}$. Now since 
$a(\tau)$, $a^{\prime}(\tau)$ and $\mu_k(\tau)$ are continuous at the transition 
surface, we immediately deduce the continuity of $\mu_k^{\prime}(\tau)$.
	
Hence we extract the matching conditions for tensor perturbations: $\mu_k$ 
and $\mu_k^{\prime}$ are continuous at the transition surface $\tau = 
\tau_{B-}$ and $\tau = \tau_{B+}$. In the present case $\mu_k$ and 
$\mu_k^{\prime}$ remain almost invariant during the bounce phase, therefore we 
can directly match $\mu_k^c(\tau_{B-})$ and $\mu_k^{\prime c}(\tau_{B-})$ with 
$\mu_k^e(\tau_{B+})$ and $\mu_k^{\prime e}(\tau_{B+})$.
	
	At $\tau = \tau_{B-}$ the tensor perturbations and  derivative behave  as
	\begin{equation}
		\mu_k^c(\tau_{B-}) \simeq -b_{t,2} \frac{\Gamma(\nu_c)}{\pi} \left( \frac{2}{k} \right)^{\nu_c} (\tau_- - \tau_{B-})^{\frac{1}{2} - \nu_c} ~, 
	\end{equation}
	\begin{equation}
		\frac{d\mu_k^c}{d\tau}(\tau_{B-}) \simeq b_{t,2} \frac{\Gamma(\nu_c)}{\pi} \left(\frac{1}{2} - \nu_c \right) \left( \frac{2}{k} \right)^{\nu_c} (\tau_- - \tau_{B-})^{-\frac{1}{2} - \nu_c} ~,
	\end{equation}
	similarly to the scalar case. At the transition surface $\tau = \tau_{B+}$ 
we have
	\begin{equation}
		\mu_k^e (\tau_{B+}) = b_{t,5} \sqrt{\tau_{B+} - \tau_+} + \frac{2}{\pi} \left[ \ln \frac{k(\tau_{B+} - \tau_+)}{2} + \gamma_E \right] b_{t,6} \sqrt{\tau_{B+} - \tau_+} ~,
	\end{equation}
	\begin{equation}
		\frac{d\mu_k^e}{d\tau} (\tau_{B+}) = \frac{1}{2\sqrt{\tau_{B+} - \tau_+}} \left[ b_{t,5} + \frac{2b_{t,6}}{\pi} \left( 2 + \gamma_E + \ln \frac{k(\tau_{B+} - \tau_+)}{2} \right) \right] ~.
	\end{equation}
	Using $\mu_k^c(\tau_{B-}) = \mu_k^e (\tau_{B+})$, 
$\frac{d\mu_k^c}{d\tau}(\tau_{B-}) = \frac{d\mu_k^e}{d\tau} (\tau_{B+})$, we 
acquire
	\begin{equation}
		b_{t,6} = \frac{\Gamma(\nu_c) 2^{\nu_c - 2}}{k^{\nu_c} \sqrt{\tau_{B+} - 
\tau_+} (\tau_-  - \tau_{B-})^{\frac{1}{2} + \nu_c}} \left[ (1-2\nu_c) 
(\tau_{B+} - \tau_+) + (\tau_- - \tau_{B-}) \right] b_{t,2} ~.
	\end{equation}
	We can simplify this relation by using
	\begin{equation}
		\tau_{B+} - \tau_+ = \frac{1}{2\mathcal{H}_{B+}} ~,~ \tau_- - \tau_{B-} = -\frac{q}{(1-q) \mathcal{H}_{B-}} ~,~ \mathcal{H}_{B-} \simeq -\mathcal{H}_{B+} ~,
	\end{equation}
	obtaining
	\begin{equation}
		\label{eq:bt6final}
		\frac{b_{t,6}}{b_{t,2}} = \frac{\left(1-2 \nu_{c}\right)^{\frac{1}{2}-\nu_{c}} \Gamma\left(\nu_{c}\right)}{4^{1-\nu_{c}}} \sqrt{\frac{\mathcal{H}_{B+}}{-\mathcal{H}_{B-}}}\left(1+\frac{-\mathcal{H}_{B-}}{\mathcal{H}_{B+}}\right)\left(\frac{-\mathcal{H}_{B-}}{k}\right)^{\nu_{c}}  b_{t,2} ~,
	\end{equation}
	\begin{equation}
		\label{eq:bt5final}
		\frac{b_{t,5}}{b_{t,2}} = \frac{-\left(1-2 \nu_{c}\right)^{\frac{1}{2}-\nu_{c}} \Gamma\left(\nu_{c}\right)}{2^{1-2 \nu_{c}} \pi} \sqrt{\frac{\mathcal{H}_{B+}}{-\mathcal{H}_{B-}}}\left[2+\left(1+\frac{-\mathcal{H}_{B-}}{\mathcal{H}_{B+}}\right)\ln \frac{ke^{\gamma_E}}{4 \mathcal{H}_{B+}} \right]\left(\frac{-\mathcal{H}_{B-}}{k}\right)^{\nu_{c}} ~.
	\end{equation}	
	Expressions \eqref{eq:bt6final} and \eqref{eq:bt5final} are the final 
results of this Appendix.



\begin{thebibliography}{99}
		
		\bibitem{Novello:2008ra} 
		M.~Novello and S.~E.~P.~Bergliaffa,
		{\it{Bouncing Cosmologies}},
		Phys.\ Rept.\  {\bf 463}, 127 (2008)
		[arXiv:0802.1634 [astro-ph]].
		
 
  
		\bibitem{Lehners:2008vx} 
		J.~L.~Lehners,
		{\it{Ekpyrotic and Cyclic Cosmology}},
		Phys.\ Rept.\  {\bf 465}, 223 (2008)
		[arXiv:0806.1245 [astro-ph]].
		
		\bibitem{Cai:2014bea} 
		Y.~F.~Cai,
		{\it{Exploring Bouncing Cosmologies with Cosmological Surveys}},
		Sci.\ China Phys.\ Mech.\ Astron.\  {\bf 57}, 1414 (2014)
		[arXiv:1405.1369 [hep-th]].
		
		\bibitem{Battefeld:2014uga} 
		D.~Battefeld and P.~Peter,
		{\it{A Critical Review of Classical Bouncing Cosmologies}},
		Phys.\ Rept.\  {\bf 571}, 1 (2015)
		[arXiv:1406.2790 [astro-ph.CO]].
		
		\bibitem{Brandenberger:2016vhg} 
		R.~Brandenberger and P.~Peter,
		{\it{Bouncing Cosmologies: Progress and Problems}},
		Found.\ Phys.\  {\bf 47}, no. 6, 797 (2017)
		[arXiv:1603.05834 [hep-th]].
		
		\bibitem{Cai:2016hea} 
		Y.~F.~Cai, A.~Marciano, D.~G.~Wang and E.~Wilson-Ewing,
		{\it{Bouncing cosmologies with dark matter and dark energy}},
		Universe {\bf 3}, no. 1, 1 (2016)
		[arXiv:1610.00938 [astro-ph.CO]].
		
		\bibitem{Brandenberger:2011gk}
		R.~H.~Brandenberger,
		{\it{Introduction to Early Universe Cosmology}},
		PoS \textbf{ICFI2010}, 001 (2010)
		[arXiv:1103.2271 [astro-ph.CO]].
		
		\bibitem{Borde:1993xh} 
		A.~Borde and A.~Vilenkin,
		{\it{Eternal inflation and the initial singularity}},
		Phys.\ Rev.\ Lett.\  {\bf 72}, 3305 (1994)
		[gr-qc/9312022].
		
		\bibitem{Borde:2001nh} 
		A.~Borde, A.~H.~Guth and A.~Vilenkin,
		{\it{Inflationary space-times are incompletein past directions}},
		Phys.\ Rev.\ Lett.\  {\bf 90}, 151301 (2003)
		[gr-qc/0110012].
		
		\bibitem{Cline:2003gs} 
		J.~M.~Cline, S.~Jeon and G.~D.~Moore,
		{\it{The Phantom menaced: Constraints on low-energy effective ghosts}},
		Phys.\ Rev.\ D {\bf 70}, 043543 (2004)
		[hep-ph/0311312].
		
		\bibitem{Easson:2016klq}
		D.~A.~Easson and A.~Vikman,
		{\it{The Phantom of the New Oscillatory Cosmological Phase}},
		[arXiv:1607.00996 [gr-qc]].
		
		\bibitem{Xia:2007km}
		J.~Q.~Xia, Y.~F.~Cai, T.~T.~Qiu, G.~B.~Zhao and X.~Zhang,
		{\it{Constraints on the Sound Speed of Dynamical Dark Energy}},
		Int. J. Mod. Phys. D \textbf{17}, 1229-1243 (2008)
		[arXiv:astro-ph/0703202 [astro-ph]].
		
		\bibitem{Vikman:2004dc} 
		A.~Vikman,
		{\it{Can dark energy evolve to the phantom?}},
		Phys.\ Rev.\ D {\bf 71}, 023515 (2005)
		[astro-ph/0407107].
		
		\bibitem{Karouby:2010wt} 
		J.~Karouby and R.~Brandenberger,
		{\it{A Radiation Bounce from the Lee-Wick Construction?}},
		Phys.\ Rev.\ D {\bf 82}, 063532 (2010)
		[arXiv:1004.4947 [hep-th]].
		
		\bibitem{Karouby:2011wj} 
		J.~Karouby, T.~Qiu and R.~Brandenberger,
		{\it{On the Instability of the Lee-Wick Bounce}},
		Phys.\ Rev.\ D {\bf 84}, 043505 (2011)
		[arXiv:1104.3193 [hep-th]].
		
		\bibitem{Bhattacharya:2013ut} 
		K.~Bhattacharya, Y.~F.~Cai and S.~Das,
		{\it{Lee-Wick radiation induced bouncing universe models}},
		Phys.\ Rev.\ D {\bf 87}, no. 8, 083511 (2013)
		[arXiv:1301.0661 [hep-th]].
		
		\bibitem{Cai:2013vm} 
		Y.~F.~Cai, R.~Brandenberger and P.~Peter,
		{\it{Anisotropy in a Nonsingular Bounce}},
		Class.\ Quant.\ Grav.\  {\bf 30}, 075019 (2013)
		[arXiv:1301.4703 [gr-qc]].
		
		\bibitem{Belinsky:1970ew} 
		V.~A.~Belinsky, I.~M.~Khalatnikov and E.~M.~Lifshitz,
		{\it{Oscillatory approach to a singular point in the relativistic cosmology}},
		Adv.\ Phys.\  {\bf 19}, 525 (1970).
		
		\bibitem{Cai:2009fn} 
		Y.~F.~Cai, W.~Xue, R.~Brandenberger and X.~Zhang,
		{\it{Non-Gaussianity in a Matter Bounce}},
		JCAP {\bf 0905}, 011 (2009)
		[arXiv:0903.0631 [astro-ph.CO]].
		
		\bibitem{Gao:2014hea} 
		X.~Gao, M.~Lilley and P.~Peter,
		{\it{Production of non-gaussianities through a positive spatial curvature bouncing phase}},
		JCAP {\bf 1407}, 010 (2014)
		[arXiv:1403.7958 [gr-qc]].
		
		\bibitem{Gao:2014eaa} 
		X.~Gao, M.~Lilley and P.~Peter,
		{\it{Non-Gaussianity excess problem in classical bouncing cosmologies}},
		Phys.\ Rev.\ D {\bf 91}, no. 2, 023516 (2015)
		[arXiv:1406.4119 [gr-qc]].
		
		\bibitem{Quintin:2015rta} 
		J.~Quintin, Z.~Sherkatghanad, Y.~F.~Cai and R.~H.~Brandenberger,
		{\it{Evolution of cosmological perturbations and the production of non-Gaussianities through a nonsingular bounce: Indications for a no-go theorem in single field matter bounce cosmologies}},
		Phys.\ Rev.\ D {\bf 92}, no. 6, 063532 (2015)
		[arXiv:1508.04141 [hep-th]].
		
		\bibitem{Li:2016xjb} 
		Y.~B.~Li, J.~Quintin, D.~G.~Wang and Y.~F.~Cai,
		{\it{Matter bounce cosmology with a generalized single field: non-Gaussianity and an extended no-go theorem}},
		JCAP {\bf 1703}, 031 (2017)
		[arXiv:1612.02036 [hep-th]].
		
		\bibitem{Akama:2019qeh} 
		S.~Akama, S.~Hirano and T.~Kobayashi,
		{\it{Primordial non-Gaussianities of scalar and tensor perturbations in general bounce cosmology: Evading the no-go theorem}},
		Phys. Rev. D \textbf{101}, no.4, 043529 (2020)
		[arXiv:1908.10663 [gr-qc]].
		
    \bibitem{Kumar:2021mgc}
    K.~S.~Kumar, S.~Maheshwari, A.~Mazumdar and J.~Peng,
    {\it{An anisotropic bouncing universe in non-local gravity}},
    JCAP \textbf{07} (2021), 025
    [arXiv:2103.13980 [gr-qc]].
		
    \bibitem{Pinto-Neto:2021gcl}
    N.~Pinto-Neto,
    {\it{Bouncing Quantum Cosmology}},
    Universe \textbf{7} (2021) no.4, 110
    
        \bibitem{Nandi:2020szp}
        D.~Nandi,
        {\it{Stability of a viable non-minimal bounce}},
        Universe \textbf{7}, no.3, 62 (2021)
        [arXiv:2009.03134 [gr-qc]].

		\bibitem{Cai:2016thi}
		Y.~Cai, Y.~Wan, H.~G.~Li, T.~Qiu and Y.~S.~Piao,
		{\it{The Effective Field Theory of nonsingular cosmology}},
		JHEP \textbf{01}, 090 (2017)
		[arXiv:1610.03400 [gr-qc]].
		
		\bibitem{Creminelli:2016zwa}
		P.~Creminelli, D.~Pirtskhalava, L.~Santoni and E.~Trincherini,
		{\it{Stability of Geodesically Complete Cosmologies}},
		JCAP \textbf{11}, 047 (2016)
		[arXiv:1610.04207 [hep-th]].
		
		\bibitem{Cai:2017tku}
		Y.~Cai, H.~G.~Li, T.~Qiu and Y.~S.~Piao,
		{\it{The Effective Field Theory of nonsingular cosmology: II}},
		Eur. Phys. J. C \textbf{77}, no.6, 369 (2017)
		[arXiv:1701.04330 [gr-qc]].
		
		\bibitem{Cai:2017pga}
Y.~Cai, Y.~T.~Wang, J.~Y.~Zhao and Y.~S.~Piao,
{\it{Primordial perturbations with pre-inflationary bounce}},
Phys. Rev. D \textbf{97}, no.10, 103535 (2018)
[arXiv:1709.07464 [astro-ph.CO]].

		\bibitem{Kolevatov:2017voe}
		R.~Kolevatov, S.~Mironov, N.~Sukhov and V.~Volkova,
		{\it{Cosmological bounce and Genesis beyond Horndeski}},
		JCAP \textbf{08}, 038 (2017)
		[arXiv:1705.06626 [hep-th]].
		
		\bibitem{Cai:2017dyi}
		Y.~Cai and Y.~S.~Piao,
		{\it{A covariant Lagrangian for stable nonsingular bounce}},
		JHEP \textbf{09}, 027 (2017)
		[arXiv:1705.03401 [gr-qc]].
		
		\bibitem{Ye:2019frg}
		G.~Ye and Y.~S.~Piao,
		{\it{Implication of GW170817 for cosmological bounces}},
		Commun. Theor. Phys. \textbf{71}, no.4, 427 (2019)
		[arXiv:1901.02202 [gr-qc]].
		
		\bibitem{Ye:2019sth}
		G.~Ye and Y.~S.~Piao,
		{\it{Bounce in general relativity and higher-order derivative operators}},
		Phys. Rev. D \textbf{99}, no.8, 084019 (2019)
		[arXiv:1901.08283 [gr-qc]].
		
    \bibitem{Gungor:2020fce}
    \"O.~G\"ung\"or and G.~D.~Starkman,
    {\it{A classical, non-singular, bouncing universe}},
    JCAP \textbf{04} (2021), 003
    [arXiv:2011.05133 [gr-qc]].
		
		\bibitem{Zheng:2017qfs}
Y.~Zheng, L.~Shen, Y.~Mou and M.~Li,
    {\it{On (in)stabilities of perturbations in mimetic models with higher derivatives}},
JCAP \textbf{08}, 040 (2017)
[arXiv:1704.06834 [gr-qc]].
		
		\bibitem{Libanov:2016kfc}
		M.~Libanov, S.~Mironov and V.~Rubakov,
		{\it{Generalized Galileons: instabilities of bouncing and Genesis cosmologies and modified Genesis}},
		JCAP \textbf{08}, 037 (2016)
		[arXiv:1605.05992 [hep-th]].
		
		\bibitem{Kobayashi:2016xpl}
		T.~Kobayashi,
		{\it{Generic instabilities of nonsingular cosmologies in Horndeski theory: A no-go theorem}},
		Phys. Rev. D \textbf{94}, no.4, 043511 (2016)
		[arXiv:1606.05831 [hep-th]].
		
		\bibitem{Akama:2017jsa}
		S.~Akama and T.~Kobayashi,
		{\it{Generalized multi-Galileons, covariantized new terms, and the no-go theorem for nonsingular cosmologies}},
		Phys. Rev. D \textbf{95}, no.6, 064011 (2017)
		[arXiv:1701.02926 [hep-th]].
		
		\bibitem{Khoury:2001wf}
		J.~Khoury, B.~A.~Ovrut, P.~J.~Steinhardt and N.~Turok,
		{\it{The Ekpyrotic universe: Colliding branes and the origin of the hot big bang}},
		Phys. Rev. D \textbf{64}, 123522 (2001)
		[arXiv:hep-th/0103239 [hep-th]].
		
		\bibitem{Ilyas:2020qja}
		A.~Ilyas, M.~Zhu, Y.~Zheng, Y.~F.~Cai and E.~N.~Saridakis,
		{\it{DHOST Bounce}},
		JCAP \textbf{09}, 002 (2020)
		[arXiv:2002.08269 [gr-qc]].
		
		\bibitem{Nicolis:2008in}
		A.~Nicolis, R.~Rattazzi and E.~Trincherini,
		{\it{The Galileon as a local modification of gravity}},
		Phys. Rev. D \textbf{79}, 064036 (2009)
		[arXiv:0811.2197 [hep-th]].
		
		\bibitem{Deffayet:2011gz}
		C.~Deffayet, X.~Gao, D.~A.~Steer and G.~Zahariade,
		{\it{From k-essence to generalised Galileons}},
		Phys. Rev. D \textbf{84}, 064039 (2011)
		[arXiv:1103.3260 [hep-th]].
		
		\bibitem{Kobayashi:2011nu}
		T.~Kobayashi, M.~Yamaguchi and J.~Yokoyama,
		{\it{Generalized G-inflation: Inflation with the most general second-order field equations}},
		Prog. Theor. Phys. \textbf{126}, 511-529 (2011)
		[arXiv:1105.5723 [hep-th]].
		
		\bibitem{Horndeski:1974wa}
		G.~W.~Horndeski,
		{\it{Second-order scalar-tensor field equations in a four-dimensional space}},
		Int. J. Theor. Phys. \textbf{10}, 363-384 (1974)
		
		\bibitem{Kobayashi:2010cm}
		T.~Kobayashi, M.~Yamaguchi and J.~Yokoyama,
		{\it{G-inflation: Inflation driven by the Galileon field}},
		Phys. Rev. Lett. \textbf{105}, 231302 (2010)
		[arXiv:1008.0603 [hep-th]].
		
		\bibitem{Deffayet:2010qz}
		C.~Deffayet, O.~Pujolas, I.~Sawicki and A.~Vikman,
		{\it{Imperfect Dark Energy from Kinetic Gravity Braiding}},
		JCAP \textbf{10}, 026 (2010)
		[arXiv:1008.0048 [hep-th]].
		
		\bibitem{Qiu:2011cy}
		T.~Qiu, J.~Evslin, Y.~F.~Cai, M.~Li and X.~Zhang,
		{\it{Bouncing Galileon Cosmologies}},
		JCAP \textbf{10}, 036 (2011)
		[arXiv:1108.0593 [hep-th]].
		
		\bibitem{Easson:2011zy}
		D.~A.~Easson, I.~Sawicki and A.~Vikman,
		{\it{G-Bounce}},
		JCAP \textbf{11}, 021 (2011)
		[arXiv:1109.1047 [hep-th]].
		
		\bibitem{Cai:2012va}
		Y.~F.~Cai, D.~A.~Easson and R.~Brandenberger,
		{\it{Towards a Nonsingular Bouncing Cosmology}},
		JCAP \textbf{08}, 020 (2012)
		[arXiv:1206.2382 [hep-th]].
		
\bibitem{Leon:2012mt}
G.~Leon and E.~N.~Saridakis,
{\it{Dynamical analysis of generalized Galileon cosmology}},
JCAP \textbf{03}, 025 (2013)
[arXiv:1211.3088 [astro-ph.CO]].

\bibitem{CANTATA:2021ktz}
E.~N.~Saridakis \textit{et al.} [CANTATA],
 {\it{Modified Gravity and Cosmology: An Update by the CANTATA Network}},
[arXiv:2105.12582 [gr-qc]].

		\bibitem{Ostrogradsky:1850fid}
		M.~Ostrogradsky,
		{\it{M\'emoires sur les \'equations diff\'erentielles, relatives au probl\`eme des isop\'erim\`etres}},
		Mem. Acad. St. Petersbourg \textbf{6}, no.4, 385-517 (1850)
		
\bibitem{An:2020lkg}
O.~S.~An, J.~U.~Kang, T.~H.~Kim and U.~R.~Mun,
{\it{Notes on the post-bounce background dynamics in bouncing cosmologies}},
[arXiv:2010.13287 [gr-qc]].
	
		\bibitem{Brandenberger:2017pjz}
		R.~Brandenberger, Q.~Liang, R.~O.~Ramos and S.~Zhou,
		{\it{Fluctuations through a Vibrating Bounce}},
		Phys. Rev. D \textbf{97}, no.4, 043504 (2018)
		[arXiv:1711.08370 [hep-th]].
		
        \bibitem{Mironov:2020mfo}
        S.~Mironov, V.~Rubakov and V.~Volkova,
        {\it{Superluminality in beyond Horndeski theory with extra scalar field}},
        Phys. Scripta \textbf{95}, no.8, 084002 (2020)
        [arXiv:2005.12626 [hep-th]].
		
        \bibitem{Mironov:2020pqh}
        S.~Mironov, V.~Rubakov and V.~Volkova,
        {\it{Superluminality in DHOST theory with extra scalar}},
        JHEP \textbf{04}, 035 (2021)
        [arXiv:2011.14912 [hep-th]].
		
        \bibitem{Jonas:2021xkx}
        C.~Jonas, J.~L.~Lehners and J.~Quintin,
        {\it{Cosmological consequences of a principle of finite amplitudes}},
        Phys. Rev. D \textbf{103}, no.10, 103525 (2021)
        [arXiv:2102.05550 [hep-th]].
		
		\bibitem{Cai:2007zv}
		Y.~F.~Cai, T.~Qiu, R.~Brandenberger, Y.~S.~Piao and X.~Zhang,
		{\it{On Perturbations of Quintom Bounce}},
		JCAP \textbf{03}, 013 (2008)
		[arXiv:0711.2187 [hep-th]].
		
		\bibitem{Cai:2008ed}
		Y.~F.~Cai and X.~Zhang,
		{\it{Evolution of Metric Perturbations in Quintom Bounce model}},
		JCAP \textbf{06}, 003 (2009)
		[arXiv:0808.2551 [astro-ph]].
		
		\bibitem{Cai:2008qw}
		Y.~F.~Cai, T.~t.~Qiu, R.~Brandenberger and X.~m.~Zhang,
		{\it{A Nonsingular Cosmology with a Scale-Invariant Spectrum of Cosmological Perturbations from Lee-Wick Theory}},
		Phys. Rev. D \textbf{80}, 023511 (2009)
		[arXiv:0810.4677 [hep-th]].
		
		\bibitem{Lin:2010pf}
		C.~Lin, R.~H.~Brandenberger and L.~Perreault Levasseur,
		{\it{A Matter Bounce By Means of Ghost Condensation}},
		JCAP \textbf{04}, 019 (2011)
		[arXiv:1007.2654 [hep-th]].
		
		\bibitem{Ilyas:2020zcb}
A.~Ilyas, M.~Zhu, Y.~Zheng and Y.~F.~Cai,
{\it{Emergent Universe and Genesis from the DHOST Cosmology}},
JHEP \textbf{01}, 141 (2021)
[arXiv:2009.10351 [gr-qc]].

\bibitem{Zhu:2021ggm}
M.~Zhu and Y.~Zheng,
{\it{Improved DHOST Genesis}},
[arXiv:2109.05277 [gr-qc]].

		
		\bibitem{Sasaki:1983kd}
		M.~Sasaki,
		{\it{Gauge Invariant Scalar Perturbations in the New Inflationary Universe}},
		Prog. Theor. Phys. \textbf{70}, 394 (1983)
		
		\bibitem{Kodama:1985bj}
		H.~Kodama and M.~Sasaki,
		{\it{Cosmological Perturbation Theory}},
		Prog. Theor. Phys. Suppl. \textbf{78}, 1-166 (1984)
		
		\bibitem{Mukhanov:1988jd}
		V.~F.~Mukhanov,
		{\it{Quantum Theory of Gauge Invariant Cosmological Perturbations}},
		Sov. Phys. JETP \textbf{67}, 1297-1302 (1988)
		
		\bibitem{Cai:2014xxa}
		Y.~F.~Cai, J.~Quintin, E.~N.~Saridakis and E.~Wilson-Ewing,
		{\it{Nonsingular bouncing cosmologies in light of BICEP2}},
		JCAP \textbf{07}, 033 (2014)
		[arXiv:1404.4364 [astro-ph.CO]].
		
		
\bibitem{Banerjee:2016hom}
S.~Banerjee and E.~N.~Saridakis,
{\it{Bounce and cyclic cosmology in weakly broken galileon theories}},
Phys. Rev. D \textbf{95}, no.6, 063523 (2017)
[arXiv:1604.06932 [gr-qc]].
		
		\bibitem{Hwang:1991an}
		J.~c.~Hwang and E.~T.~Vishniac,
		{\it{Gauge-invariant joining conditions for cosmological perturbations}},
		Astrophys. J. \textbf{382}, 363-368 (1991)
		
		\bibitem{Deruelle:1995kd}
		N.~Deruelle and V.~F.~Mukhanov,
		{\it{On matching conditions for cosmological perturbations}},
		Phys. Rev. D \textbf{52}, 5549-5555 (1995)
		[arXiv:gr-qc/9503050 [gr-qc]].
		
		\bibitem{Creminelli:2012my}
		P.~Creminelli, K.~Hinterbichler, J.~Khoury, A.~Nicolis and E.~Trincherini,
		{\it{Subluminal Galilean Genesis}},
		JHEP \textbf{02}, 006 (2013)
		[arXiv:1209.3768 [hep-th]].
		
		\bibitem{Hinterbichler:2012yn}
		K.~Hinterbichler, A.~Joyce, J.~Khoury and G.~E.~J.~Miller,
		{\it{Dirac-Born-Infeld Genesis: An Improved Violation of the Null Energy Condition}},
		Phys. Rev. Lett. \textbf{110}, no.24, 241303 (2013)
		[arXiv:1212.3607 [hep-th]].
		
		\bibitem{Padilla:2012ze}
		A.~Padilla and V.~Sivanesan,
		{\it{Boundary Terms and Junction Conditions for Generalized Scalar-Tensor Theories}},
		JHEP \textbf{08}, 122 (2012)
		[arXiv:1206.1258 [gr-qc]].
		
		\bibitem{Nishi:2014bsa}
		S.~Nishi, T.~Kobayashi, N.~Tanahashi and M.~Yamaguchi,
		{\it{Cosmological matching conditionsand galilean genesis in Horndeski's theory}},
		JCAP \textbf{03}, 008 (2014)
		[arXiv:1401.1045 [hep-th]].
		
		\bibitem{Aviles:2019xae}
		L.~Avil\'es, H.~Maeda and C.~Martinez,
		{\it{Junction conditions in scalar\textendash{}tensor theories}},
		Class. Quant. Grav. \textbf{37}, no.7, 075022 (2020)
		[arXiv:1910.07534 [gr-qc]].
		
\bibitem{Saridakis:2021qxb}
E.~N.~Saridakis,
 {\it{Do we need soft cosmology?}},
[arXiv:2105.08646 [astro-ph.CO]].

		\bibitem{Lyth:2001pf}
		D.~H.~Lyth,
	{\it{The Primordial curvature perturbation in the ekpyrotic universe}},
		Phys. Lett. B \textbf{524}, 1-4 (2002)
		[arXiv:hep-ph/0106153 [hep-ph]].
		
		\bibitem{Brandenberger:2001bs}
		R.~Brandenberger and F.~Finelli,
		{\it{On the spectrum of fluctuations in an effective field theory of the Ekpyrotic universe}},
		JHEP \textbf{11}, 056 (2001)
		[arXiv:hep-th/0109004 [hep-th]].
		
		\bibitem{Tsujikawa:2002qc}
		S.~Tsujikawa, R.~Brandenberger and F.~Finelli,
		{\it{On the construction of nonsingular pre - big bang and ekpyrotic cosmologies and the resulting density perturbations}},
		Phys. Rev. D \textbf{66}, 083513 (2002)
		[arXiv:hep-th/0207228 [hep-th]].
		
		\bibitem{Tolley:2003nx}
		A.~J.~Tolley, N.~Turok and P.~J.~Steinhardt,
		{\it{Cosmological perturbations in a big crunch / big bang space-time}},
		Phys. Rev. D \textbf{69}, 106005 (2004)
		[arXiv:hep-th/0306109 [hep-th]].
		
		\bibitem{Notari:2002yc}
		A.~Notari and A.~Riotto,
		{\it{Isocurvature perturbations in the ekpyrotic universe}},
		Nucl. Phys. B \textbf{644}, 371-382 (2002)
		[arXiv:hep-th/0205019 [hep-th]].
		
		\bibitem{Finelli:2002we}
		F.~Finelli,
		{\it{Assisted contraction}},
		Phys. Lett. B \textbf{545}, 1-7 (2002)
		[arXiv:hep-th/0206112 [hep-th]].
		
		\bibitem{Creminelli:2007aq}
		P.~Creminelli and L.~Senatore,
		{\it{A Smooth bouncing cosmology with scale invariant spectrum}},
		JCAP \textbf{11}, 010 (2007)
		[arXiv:hep-th/0702165 [hep-th]].
		
		\bibitem{Lehners:2007ac}
		J.~L.~Lehners, P.~McFadden, N.~Turok and P.~J.~Steinhardt,
		{\it{Generating ekpyrotic curvature perturbations before the big bang}},
		Phys. Rev. D \textbf{76}, 103501 (2007)
		[arXiv:hep-th/0702153 [hep-th]].
		
		\bibitem{Buchbinder:2007ad}
		E.~I.~Buchbinder, J.~Khoury and B.~A.~Ovrut,
		{\it{New Ekpyrotic cosmology}},
		Phys. Rev. D \textbf{76}, 123503 (2007)
		[arXiv:hep-th/0702154 [hep-th]].
		
		\bibitem{Battefeld:2005wv}
		T.~J.~Battefeld, S.~P.~Patil and R.~H.~Brandenberger,
		{\it{On the transfer of metric fluctuations when extra dimensions bounce or stabilize}},
		Phys. Rev. D \textbf{73}, 086002 (2006)
		[arXiv:hep-th/0509043 [hep-th]].
		\bibitem{Wands:1998yp}
		D.~Wands,
		{\it{Duality invariance of cosmological perturbation spectra}},
		Phys. Rev. D \textbf{60}, 023507 (1999)
		[arXiv:gr-qc/9809062 [gr-qc]].
		
		\bibitem{Durrer:2002jn}
		R.~Durrer and F.~Vernizzi,
		{\it{Adiabatic perturbations in pre - big bang models: Matching conditions and scale invariance}},
		Phys. Rev. D \textbf{66}, 083503 (2002)
		[arXiv:hep-ph/0203275 [hep-ph]].
		
		\bibitem{Cai:2013kja}
		Y.~F.~Cai, E.~McDonough, F.~Duplessis and R.~H.~Brandenberger,
		{\it{Two Field Matter Bounce Cosmology}},
		JCAP \textbf{10}, 024 (2013)
		[arXiv:1305.5259 [hep-th]].
		
        \bibitem{Aghanim:2018eyx}
        N.~Aghanim \textit{et al.} [Planck],
        {\it{Planck 2018 results. VI. Cosmological parameters}},
        Astron. Astrophys. \textbf{641}, A6 (2020)
        [arXiv:1807.06209 [astro-ph.CO]].
		
		
		\bibitem{Cai:2009rd}
		Y.~F.~Cai, W.~Xue, R.~Brandenberger and X.~m.~Zhang,
		{\it{Thermal Fluctuations and Bouncing Cosmologies}},
		JCAP \textbf{06}, 037 (2009)
		[arXiv:0903.4938 [hep-th]].
	

        \bibitem{Enqvist:2001zp}
        K.~Enqvist and M.~S.~Sloth,
        {\it{Adiabatic CMB perturbations in pre - big bang string cosmology}},
        Nucl. Phys. B \textbf{626}, 395-409 (2002)
        [arXiv:hep-ph/0109214 [hep-ph]].
		
        \bibitem{Lyth:2001nq}
        D.~H.~Lyth and D.~Wands,
        {\it{Generating the curvature perturbation without an inflaton}},
        Phys. Lett. B \textbf{524}, 5-14 (2002)
        [arXiv:hep-ph/0110002 [hep-ph]].
			
		\bibitem{Cai:2011zx}
		Y.~F.~Cai, R.~Brandenberger and X.~Zhang,
		{\it{The Matter Bounce Curvaton Scenario}},
		JCAP \textbf{03}, 003 (2011)
		[arXiv:1101.0822 [hep-th]].
		
		\bibitem{Cai:2011ci}
		Y.~F.~Cai, R.~Brandenberger and X.~Zhang,
		{\it{Preheating a bouncing universe}},
		Phys. Lett. B \textbf{703}, 25-33 (2011)
		[arXiv:1105.4286 [hep-th]].
	  
        \bibitem{Tahara:2020fmn}
        H.~W.~H.~Tahara and T.~Kobayashi,
        {\it{Nanohertz gravitational waves from a null-energy-condition violation in the early universe}},
        Phys. Rev. D \textbf{102}, no.12, 123533 (2020)
        [arXiv:2011.01605 [gr-qc]].
	
        \bibitem{Cai:2020qpu}
        Y.~Cai and Y.~S.~Piao,
        {\it{Intermittent null energy condition violations during inflation and primordial gravitational waves}},
        Phys. Rev. D \textbf{103}, no.8, 083521 (2021)
        [arXiv:2012.11304 [gr-qc]].
	
	\end{thebibliography}
\end{document}